\documentclass[twocolumn,pre,floatfix]{revtex4}
\usepackage{psfrag,epsfig,amsfonts,amssymb,amsmath,wasysym,bm}
\usepackage{dcolumn}

\usepackage[normalem]{ulem}
\usepackage{color}

\newcommand{\lmax}{\lambda_\mathrm{max}}
\newcommand{\lmin}{\lambda_\mathrm{min}}
\newcommand{\EF}{E_\mathrm{F}}
\newcommand{\kB}{k_\mathrm{B}}
\newcommand{\tB}{t_\mathrm{B}}
\newcommand{\lu}{\left[}
\newcommand{\ru}{\right]_{U}}

\newcommand{\rhobar}{\omega}
\newcommand{\rhou}{\rho_{\mathrm{av}}}
\newcommand{\DD}{D}
\newcommand{\hh}{\hbar}
\newcommand{\tr}{\mbox{Tr}}
\newcommand{\hr}{{\cal H}}
\newcommand{\ord}{{\cal O}}
\newcommand{\propa}{{\cal U}_t}
\newcommand{\propb}{{\cal U}'_t}

\newcommand{\rhomic}{\rho_{\mathrm{mc}}}
\newcommand{\tto}{\rightsquigarrow}
\newcommand{\da}{\Delta_{\! A}}

\begin{document}

\title{Typical fast thermalization processes in closed many-body systems}

\author{Peter Reimann}

\affiliation{Fakult\"at f\"ur Physik, Universit\"at Bielefeld, 
33615 Bielefeld, Germany}

\begin{abstract}
Lack of knowledge about the detailed
many-particle motion on the microscopic 
scale is a key issue in any theoretical 
description of a macroscopic experiment.
For systems at or close to thermal equilibrium,
statistical mechanics provides a very 
successful general framework 
to cope with this problem.
Far from equilibrium, only very few 
quantitative and comparably universal
results are known.
Here, a new quantum mechanical 
prediction of this type is derived and verified against 
various experimental and numerical data 
from the literature. 
It quantitatively 
describes the entire temporal 
relaxation towards thermal equilibrium for a 
large class (in a mathematically precisely defined 
sense) of closed many-body systems, 
whose initial state may be arbitrarily far 
from equilibrium.
\vspace*{1cm}
\end{abstract}

\maketitle

In a macroscopic object, which is spatially 
confined and unperturbed by the rest of the 
world, every single atom exhibits an 
essentially unpredictable, 
chaotic motion ad infinitum, yet the 
system as a whole seems to 
approach
in a predictable
and often relatively simple manner 
some steady equilibrium state.
Paradigmatic examples are compound 
systems, parts of which are initially 
hotter than others, or a simple gas in 
a box, streaming through a little 
hole into an empty second box.
While such equilibration and thermalization
phenomena are omnipresent in daily life and 
extensively observed in experiments,
they entail some very challenging 
fundamental questions:
Why are the macroscopic phenomena
reproducible though the microscopic details 
are irreproducible in any real experiment?
How can the irreversible tendency
towards macroscopic equilibrium 
be reconciled with the basic laws 
of physics, implying a perpetual 
and essentially reversible motion
on the microscopic level?

Such fundamental issues
are widely considered as 
still not satisfactorily understood 
\cite{tas98,pop06,gol06b,rig08,gem09,eis15}.
Within the realm of classical mechanics, 
they go back to Maxwell,
Boltzmann, and many others \cite{skl93}.
Their quantum mechanical treatment was 
initiated by von Neumann \cite{neu29}
and is presently 
attracting renewed
interest 
\cite{gol10a,gol10b,gol10c,rei15},
e.g., in the context of 
imitating 
thermal equilibrium 
by single pure states due to such fascinating 
phenomena as concentration of measure 
\cite{pop06,pop06a,mul11}, 
canonical typicality 
\cite{gol06b,sug07,rei07,bar09,sug12},
or eigenstate thermalization 
\cite{deu91,sre94,rig08,neu12,rig12,ike13,beu14,ste14,gol15c}.
Numerically, scrutinizing ultracold atom experiments 
\cite{cra08a,tro12,gri12,per14}
and unraveling the relations
between thermalization, integrability, 
and many-body localization are among 
the current key issues 
\cite{rig08,rig09b,rig09a,san10a,pal10,bri10,gog11,ban11}.
Analytically, essential equilibration and 
thermalization properties of closed 
many-body systems or of subsystems thereof 
were deduced from 
first principles
under increasingly weak assumptions 
about the initial disequilibrium, the system 
Hamiltonian, and the observables
\cite{tas98,neu29,gol10a,gol10b,gol10c,rei15,cra08b,rei08,lin09,rei10,sho11,rei12a,rei12b,sho12}.
In particular, groundbreaking
results regarding 
pertinent relaxation time scales 
have been obtained in
\cite{sho12,cra12,gol13,mon13,mal14,gol15,gol15b}.
Of foremost relevance for our present study
is the work of the Bristol collaboration 
\cite{mal14}, showing, among others, 
that all two-outcome measurements,
where one of the projectors is of low rank, 
equilibrate as fast as they possibly can 
without violating the time-energy
uncertainty relation.
A second recent key result is due to 
Goldstein, Hara, and Tasaki \cite{gol15,gol15b},
demonstrating that most systems 
closely approach an overwhelmingly large,
so-called equilibrium Hilbert-subspace
on the extremely short 
Boltzmann time scale $\tB :=h/\kB T$.
A more detailed account of
pertinent previous works 
is provided as Supplementary Note 1.

Here, we will further extend these
findings in two essential respects:
Instead of upper bounds
for some suitably defined characteristic 
time scale, as in \cite{mal14,gol15,gol15b},
the entire temporal relaxation will be 
approximated in the form of an equality.
As an even more decisive generalization of 
\cite{mal14,gol15,gol15b},
we will admit largely arbitrary 
observables.
Finally, and actually for the first time 
within the realm of the above mentioned 
analytical approaches 
\cite{tas98,neu29,gol10a,gol10b,gol10c,rei15,cra08b,rei08,lin09,rei10,sho11,rei12a,rei12b,sho12,cra12,gol13,mon13,mal14,gol15,gol15b},
we will compare our predictions
with various experimental as well 
as numerical data from the 
literature.
In fact, most of those data have not 
been quantitatively explained by 
any other analytical theory before.
Adopting a ``typicality approach'' similar 
in spirit to random matrix theory 
\cite{gol10a,gol10b,gol10c,rei15},
our result covers the vast majority 
(in a suitably defined mathematical sense) of initial 
conditions, observables, and system Hamiltonians.
On the other hand, many commonly considered
observables and initial conditions
actually seem to be rather special in that 
they are close to or governed by a 
hidden conserved quantity and therefore 
thermalize ``untypically slowly''.

\section*{RESULTS}
\subsection*{Setup}
Employing textbook quantum mechanics,
we consider time-independent Hamiltonians
$H$ with eigenvalues $E_n$ and eigenvectors 
$|n\rangle$ on a Hilbert space $\hr$ 
of large (but finite) dimensionality $\DD \gg 1$.
As usual, system states (pure or mixed) 
are described by density operators 
$\rho:\hr\to\hr$
and observables by Hermitian
operators $A:\hr\to\hr$
with matrix elements
$\rho_{mn}:=\langle m|\rho|n\rangle$
and 
$A_{mn}:=\langle m |A| n \rangle$,
respectively.
Expectation values are given by
$\langle A\rangle_{\!\rho}:=
\tr\{\rho A\}$
and the time evolution by
$\rho(t)=\propa\rho(0)\propa^\dagger$
with propagator $\propa:=e^{-iHt/\hh}$,
yielding
\begin{eqnarray}
\langle A\rangle_{\!\rho(t)}
= \sum_{m,n=1}^{\DD}   
\rho_{mn}(0) A_{nm}
\, e^{i(E_n-E_m)t/\hh} 
\ .
\label{10}
\end{eqnarray}

The main examples are 
closed many-body systems with a
macroscopically well defined energy,
i.e., all relevant eigenvalues 
$E_1$,...,$E_{\DD}$
are contained in some microcanonical 
energy window $[E-\Delta E,E]$,
where $\Delta E$ is small on 
the macroscopic but large 
on the microscopic scale.
For systems with $f \gg 1$ degrees 
of freedom, $\DD$ is then 
exponentially large in $f$ \cite{gol10a,rei10}.
Accordingly, the relevant 
Hilbert space $\hr$ is spanned by 
the eigenvectors $\{|n\rangle\}_{n=1}^{\DD}$
and is sometimes also named energy shell
or active Hilbert space, see, e.g., Refs.
\cite{neu29,gol10a,gol10b,gol10c,rei15}
and Supplementary Note 2 for more details.

\subsection*{Analytical results}
Our main players are the three
Hermitian operators $H$ (Hamiltonian), 
$A$ (observable), and $\rho(0)$ 
(initial state),
each with its own eigenvalues 
(spectrum) and eigenvectors 
(basis of $\hr$).
In the following, the three spectra 
will be considered as arbitrary but 
fixed, while the eigenbases will be 
randomly varied relatively to each other.
More precisely, all unitary
transformations $U :\hr\to\hr$ between 
the eigenbases of $H$ and $A$ are 
considered as equally likely
(Haar distributed \cite{neu29,gol10a,gol10b,gol10c}),
while the basis of $\rho(0)$ relatively
to that of $A$ is arbitrary but fixed.
(Equivalently, we could let ``rotate''
$H$ relatively to $\rho(0)$
while keeping $A$ fixed relatively 
to $\rho(0)$).
In particular, 
the initial expectation value 
$\langle A\rangle_{\!\rho(0)}$ 
can be chosen arbitrary but then 
remains fixed ($U$-independent).
It is only for times $t>0$ that 
the randomness of the unitary $U$ 
also randomizes (via $H$)
the further temporal evolution 
of $\rho(t)$ and thus of 
$\langle A\rangle_{\!\rho(t)}$.

The basic idea behind this randomization
of $U$ is akin to random matrix theory
\cite{gol10a,gol10b,gol10c,rei15}, namely
to derive an approximation for 
$\langle A\rangle_{\!\rho(t)}$
which applies to the overwhelming 
majority of all those randomly 
sampled $U$'s, hence 
it typically should apply also to 
the particular (non-random) $U$ 
of the actual system of interest.
A more detailed justification 
of this ``typicality approach''
will be provided in section 
``Typicality of thermalization''.

Since $A_{mn}$ refers to
the basis of $H$, these matrix elements
depend on $U$, and likewise for
$\rho_{mn}(0)$
(the explicit formulae are provided in
``Methods: Basic matrices'').
Indicating averages over $U$ by the
symbol $\lu\cdots\ru$ and exploiting
that all basis transformations $U$ are
equally likely, it follows for symmetry reasons 
that $\lu \rho_{nn}(0) A_{nn}\ru$
must be independent of $n$.
Likewise, $\lu \rho_{mn}(0)A_{nm}\ru$
must be independent of $m$ and $n$ for all 
$m\not = n$.
We thus can conclude that for any $n$
\begin{eqnarray}
\!\! \!\! \!\! \!\! 
\DD\, \lu \rho_{nn}(0)A_{nn}\ru
& = & 
\lu \sum_{k=1}^\DD \rho_{kk}(0)A_{kk}\ru
\label{20}
\end{eqnarray}
and that for any $m\not=n$
\begin{eqnarray} 
\!\! \!\! \!\! \!\! 
\DD(\DD \!\! & - & \!\! 1)\, \lu \rho_{mn}(0)  A_{nm}\ru 
= 
\lu \sum_{j\not=k} \rho_{jk}(0)A_{kj}\ru
\nonumber
\\
& = & 
\lu \sum_{j,k=1}^\DD \rho_{jk}(0)A_{kj}\ru
-\lu \sum_{k=1}^\DD \rho_{kk}(0)A_{kk}\ru \ .
\label{30}
\end{eqnarray}
Defining the auxiliary density operator $\rhobar$ 
via the matrix elements 
$\rhobar_{mn} := \delta_{mn} \rho_{nn}{(0)}$,
equation (\ref{20}) can be rewritten as
$\lu \tr\{\rhobar A\}\ru$.
Working in a reference frame where only
$H$ (and thus $\rhobar$) 
changes with $U$, 
but not $A$ and $\rho(0)$, implies 
$\lu \tr\{\rhobar A\}\ru
=\tr\{\lu \rhobar\ru A\}$.
With
$\rhou := \lu \rhobar \ru$
it follows that 
\begin{eqnarray}
\lu \rho_{nn}(0)A_{nn}\ru = \tr\{\rhou A\}/\DD
=\langle A\rangle_{\!\rhou}/\DD 
\label{50}
\end{eqnarray}
for arbitrary $n$.
Likewise, equation (\ref{30}) yields 
\begin{eqnarray} 
\lu \rho_{mn}(0)  A_{nm}\ru = 
\frac{\langle A\rangle_{\!\rho(0)}
      -\langle A\rangle_{\!\rhou} }{\DD(\DD - 1)}
\label{60}
\end{eqnarray}
for arbitrary $m\not = n$.

Upon separately averaging in equation (\ref{10})  
the summands with  $m=n$ and those with 
$m\not =n$ over $U$, 
and then exploiting equations (\ref{50}) and (\ref{60}) 
one readily finds that 
\begin{eqnarray}
\lu \langle A\rangle_{\!\rho(t)}\ru 
& = & 
\langle A\rangle_{\!\rhou} 
+
F(t)\, \left\{ \langle A\rangle_{\!\rho(0)} 
- \langle A\rangle_{\!\rhou}\right\}
\label{70}
\\
F(t) & := & \frac{D}{D-1}\left(|\phi(t)|^2-\frac{1}{D}\right)
\label{80}
\end{eqnarray}
where $\phi(t)$ is the Fourier transform of the 
spectral density from Ref. \cite{cra12}
(see also \cite{gol15b,zni11,mon14})
\begin{eqnarray}
\phi(t) :=\frac{1}{\DD}\sum_{n=1}^{\DD}e^{i E_n t/\hh} 
\ .
\label{90}
\end{eqnarray}

The following results can be derived in 
principle along similar lines
(symmetry arguments being 
one key ingredient),
but since the actual details are 
quite tedious, they are 
postponed to ``Methods''.
As a first result, one obtains
\begin{eqnarray}
\langle A\rangle_{\!\rhou}=
\langle A\rangle_{\!\rhomic}
+
\frac{\langle A\rangle_{\!\rho(0)}
      -\langle A\rangle_{\!\rhomic} }{\DD+1} \ ,
\label{100}
\end{eqnarray}
where $\rhomic:=I/\DD$ is the microcanonical
density operator and $I$ the identity on $\hr$.
As a second result, one finds for the statistical
fluctuations
\begin{eqnarray}
\xi(t):=
\langle A\rangle_{\!\rho(t)} 
- 
\lu \langle A\rangle_{\!\rho(t)}\ru
\label{110}
\end{eqnarray}
the estimate
\begin{eqnarray}
\lu \xi^2(t)\ru 
=
\ord (\da^2\tr\{\rho^2(0)\}/\DD)
\label{120}
\end{eqnarray}
for arbitrary $t$,
where $\da$ is the range of $A$,
i.e., the difference between the largest 
and smallest eigenvalues of $A$.
Since averaging over $U$ and integrating over
$t$ are commuting operations, equation (\ref{120}) implies that
\begin{eqnarray}
\lu \frac{1}{t_2-t_1}\int_{t_1}^{t_2} \xi^2(t)\, dt \ru
=
\ord \left(\frac{\da^2\tr\{\rho^2(0)\}}{\DD}\right)
\label{130}
\end{eqnarray}
for arbitrary $t_2>t_1$.

Considering $t$ in equation (\ref{120})
as arbitrary but fixed, equation (\ref{110})
and $D \gg 1$ imply (obviously or by 
exploiting Chebyshev's inequality  
\cite{tas98,gol10a,lin09,sho12,rei12a})
that $\langle A\rangle_{\!\rho(t)}$ 
is practically indistinguishable from 
the average in (\ref{70})
for the vast majority of all unitaries $U$.
Indeed, the fraction (normalized Haar measure)
of exceptional $U$'s is unimaginably
small for typical macroscopic systems
with, say, $f\approx 10^{23}$ degrees 
of freedom, since $\DD$ in (\ref{120}) is 
exponentially large in $f$ 
(see below equation (\ref{10})).
Likewise, considering an arbitrary but fixed 
time interval $[t_1,t_2]$ in equation (\ref{130}),
it follows for all but a tiny fraction 
of $U$'s that the time average over
$\xi^2(t)$ on the left hand side of (\ref{130})
must be unimaginably small, and hence
also the integrand $\xi^2(t)$ itself
must be exceedingly small 
for the overwhelming majority of 
all $t\in [t_1,t_2]$.
Accordingly, $\langle A\rangle_{\!\rho(t)}$
must remain extremely close to (\ref{70})
simultaneously for all those 
$t\in [t_1,t_2]$.

Due to equation (\ref{100}) and $\DD\gg 1$,
we furthermore can safely approximate 
$\langle A\rangle_{\!\rhou}$ in (\ref{70})
by 
$\langle A\rangle_{\!\rhomic}$.
Altogether, we thus can conclude that
in very good approximation
\begin{eqnarray}
\langle A\rangle_{\!\rho(t)}
& = & 
\langle A\rangle_{\!\rhomic}
+
F(t)\, \left\{\langle A\rangle_{\!\rho(0)} 
- \langle A\rangle_{\!\rhomic}\right\}
\label{140}
\end{eqnarray}
for the vast majority of 
unitaries $U$ and times $t$.
As detailed in ``Methods'',
the neglected corrections in (\ref{140})
consist of a systematic ($U$-independent) 
part, which is bounded in modulus by 
$\da/(D^2-1)$ for all $t$, and
a random ($U$-dependent) 
part (namely $\xi(t)$),
whose typical order of magnitude is
$\da\sqrt{\tr\{\rho^2(0)\}/D}$ (for most $U$ and $t$, 
cf. equations (\ref{120}), (\ref{130})),
i.e., $\xi(t)$ is dominating by far
(note that $1\geq\tr\{\rho^2(0)\}\geq\tr\{\rhomic^2\}=1/\DD$).
Moreover, the correlations of $\xi(t)$ decay on 
time scales comparable to those 
governing $F(t)$.

These are our main formal results.
In the rest of the paper we discuss
their physical content.

\subsection*{Basic properties of $F(t)$}
Equation (\ref{90}) implies that $\phi(0)=1$, $\phi(-t)=\phi^\ast (t)$,
and $|\phi(t)| \leq 1$. 
With equation (\ref{80}) and $D\gg 1$ it follows that
in very good approximation
\begin{equation}
F(t) = |\phi(t)|^2 \ ,
\label{150}
\end{equation}
and thus
\begin{eqnarray}
F(0)=1 
\, , \
0\leq F(t)\leq 1 
\, , \
F(-t)=F(t) \ .
\label{160}
\end{eqnarray}

Indicating averages over all $t \geq 0$ by an 
overbar, one can infer from equations (\ref{90}) 
and (\ref{150}) that $\overline{F(t)}=\sum_k d_k^2/\DD^2$, 
where $k$ labels the eigenspaces of $H$ with mutually 
different eigenvalues and $d_k$ denotes their dimensions.
Since $\sum_k d_k=\DD$ we thus obtain
$\overline{F(t)} \leq \max_k(d_k/\DD)$.
Excluding extremely large multiplicities 
(degeneracies) of energy eigenvalues,
it follows that the time average
$\overline{F(t)}$ is negligibly small
and hence
\cite{tas98,gol10a,lin09,sho12,rei12a}
that $F(t)$ itself must be
negligibly small for the overwhelming majority 
of all sufficiently large $t$,
symbolically indicated as
\begin{eqnarray}
F(t\to\infty )\tto 0 \ .
\label{170}
\end{eqnarray}
Note that there still exist arbitrarily large exceptional 
$t$'s owing to the quasi-periodicity of 
$\phi(t)$ implied by (\ref{90}).
We also emphasize that our main result 
(\ref{140}) itself admits arbitrary 
degeneracies of $H$.

As an example, we focus on the
microcanonical setup introduced 
below equation (\ref{10}) and 
on not too large times, so that
(\ref{90}) is well approximated by 
\begin{eqnarray}
\phi(t)=\int_{E-\Delta E}^{E}  \rho(x)\, e^{ixt/\hh}\, dx \ ,
\label{180}
\end{eqnarray}
where $\rho(x)$ represents the (smoothened and normalized) 
density of energy levels $E_n$ in the vicinity of
the reference energy $x$.
If the level density is constant throughout
the energy window $[E-\Delta E,E]$,
we thus obtain with (\ref{150}) 
\begin{eqnarray}
F(t)=\frac{\sin^2(\Delta E\,t/2\hh)}{(\Delta E\, t/2\hh)^2} \ .
\label{190}
\end{eqnarray}

Next, we recall Boltzmann's entropy 
formula $S(x)=\kB \ln(\Omega(x))$,
where $\Omega(x)$ counts the number of 
$E_n$'s below $x$ and $\kB $ is 
Boltzmann's constant.
Hence, $\Omega'(x)$ 
must be proportional to
the level density $\rho(x)$ from above.
Furthermore,  $T:=1/S'(E)$ is the usual 
microcanonical temperature of a system 
with energy $E$ at thermal equilibrium.
A straightforward expansion then yields the
approximation $\rho(E-y)=c\, e^{-y/\kB T}$
for $y \geq 0$, where $c$ is fixed via
$\int_{E-\Delta E}^{E} \rho(x)\, dx=1$.
The omitted higher order terms 
are safely negligible for all $y\geq 0$
and systems with $f\gg 1$
degrees of freedom, see also \cite{mon14}.
With equations (\ref{150}) and (\ref{180}) one thus finds
\begin{eqnarray}
F(t)=
\frac{1-2\alpha\cos(\Delta E\, t/\hbar)+\alpha^2}
{(1-\alpha)^2[1+(\kB T\, t/\hh)^2]} \ ,
\label{200}
\end{eqnarray}
where $\alpha:=e^{-\Delta E/\kB T}$.
For $\Delta E\ll \kB T$, one recovers
(\ref{190}) and for $\Delta E\gg \kB T$
one obtains
\begin{eqnarray}
F(t)=\frac{1}{1+(\kB T\, t/\hh)^2} \ .
\label{210}
\end{eqnarray}

\subsection*{Typicality of thermalization}

Equations (\ref{140}) and (\ref{170})
imply thermalization
in the sense that the expectation value
$\langle A\rangle_{\!\rho(t)}$ becomes
(for most $U$)
practically indistinguishable from
the microcanonical average 
$\langle A\rangle_{\!\rhomic}$ for the
overwhelming majority of all sufficiently 
large $t$.
Exceptional $t$'s are, for instance,
due to quantum revivals,
which, in turn, are apparently closely
related to the quasi-periodicities 
of $F(t)$.

Our assumption that energy eigenvalues must not be
extremely highly degenerate
(see above equation (\ref{170}))
is similar to Refs. \cite{cra12,gol13,mal14,gol15,gol15b}
but considerably weaker than the corresponding 
premises in most other related works 
\cite{tas98,neu29,rei08,gol10a,gol10b,gol10c,lin09,rei10,sho11,sho12,rei12a,rei12b,rei15}.

The usual time inversion invariance
on the fundamental, microscopic level \cite{skl93}
is maintained by (\ref{140}) due to (\ref{160}).
Surprisingly, and in accordance with the second law of 
thermodynamics, the latter symmetry persists even
if it is broken in the microscopic quantum dynamics, 
e.g., by an external magnetic field!

By propagating $\rho(0)$ backward in time 
(with respect to one particular $U$) and 
taking the result as new initial state,
one may easily tailor \cite{rei10}
examples of the very rare 
$U$'s and $t$'s which notably
deviate from the typical behavior
(\ref{140}).
Equivalently, one may 
back-propagate $A$ instead of 
$\rho(0)$ (Heisenberg picture).

Note that $S$ and $T$ were 
introduced below equation (\ref{190}) not 
in the sense of associating 
some entropy and temperature to the non-equilibrium 
states $\rho(t)$, but rather as a convenient 
level-counting tool.
However, we now can identify them {\em a posteriori}
with the pertinent
entropy and temperature
after thermalization.

The randomization via $U$
(see section ``Analytical results'')
can be viewed in two ways:
Either one considers $\rho(0)$, 
$A$, and the spectrum of
$H$ as arbitrary but fixed, while the eigenbasis
of $H$ is sampled from a uniform distribution
(Haar measure).
Or one considers $H$ and the spectra of 
$\rho(0)$ and $A$ as arbitrary but fixed
and randomizes the eigenvectors of $A$ and $\rho(0)$.
In doing so, a key point is that the relative 
orientation of the eigenbases of $\rho(0)$ 
and $A$ can be chosen arbitrarily but 
then is kept fixed.
Indeed, it is well known \cite{mal14,rei15}
that for ``most'' such orientations
the expectation values $\langle A\rangle_{\!\rho(0)}$
and $\langle A\rangle_{\!\rhomic}$
are practically indistinguishable, i.e., 
an initial $\langle A\rangle_{\!\rho(0)}$ 
far from equilibrium requires
a careful fine-tuning of 
$\rho(0)$ relatively to $A$.

In reality, there is usually nothing random in the
actual physical systems one has in mind.
Hence, results like (\ref{140}), which 
(approximately) apply to the overwhelming 
majority of unitaries $U$, should be physically 
interpreted according to the common 
lore of random matrix theory
\cite{gol10a,gol10b,rei15}, 
namely as to apply practically for sure
to a concrete system under consideration,
unless there are particular reasons 
to the contrary.

Such reasons arise, for instance, 
when $A$ is known to be a 
conserved quantity,
implying a common eigenbasis of 
$A$ and $H$, i.e., the basis 
transformations $U$ must indeed
be very special.
Furthermore,
this non-typicality is structurally
stable against sufficiently
small perturbations of $A$ and/or $H$
so that the eigenvectors remain 
``almost aligned'' (each eigenvector of 
$A$ mainly overlaps with one or a few
eigenvectors of $H$) and hence
$A$ remains ``almost conserved''
(almost commuting with $H$).
Analogous non-typical $U$'s are expected
when $\rho(0)$ is known to be (almost) 
conserved (commuting with $H$).

Further well-known exceptions are 
integrable systems, 
for which thermalization in the 
above sense may be absent for
certain $\rho(0)$ and $A$
\cite{rig08,rig09a} 
(but not for others 
\cite{rig12}),
systems exhibiting many-body localization 
\cite{pal10,gog11},
or trivial cases with
non-interacting subsystems
(see also Supplementary Note 2).

Our present focus is different:
Taking thermalization for granted,
is the temporal relaxation well 
approximated by equation 
(\ref{140})?

\subsection*{Typical fast relaxation and prethermalization}
Equation (\ref{210}) is governed by the Boltzmann 
time $\tB :=h/\kB  T$, amounting 
to $\tB \approx 10^{-13}\,$s at room 
temperature.
Equation (\ref{200}) gives rise to
comparably short time scales, 
unless the temperature is exceedingly 
low or the energy window $\Delta E$ 
is unusually small. 
Such relaxation times are much shorter than
commonly observed in real systems
\cite{cra12,mal14,gol15,gol15b}.
Moreover, the temporal decay is typically
non-exponential (see e.g. 
(\ref{190})-(\ref{210})), 
again in contrast to the usual 
findings.

This seems to imply that typical 
experiments correspond 
to non-typical unitaries $U$.
Plausible explanations are as follows:
To begin with, the above predicted typical
relaxation times are so short that they
simply could not be observed in most 
experiments.
Second (or as a consequence), 
the usual initial conditions and/or
observables are indeed quite 
``special'' with respect to the 
prominent role of almost conserved 
quantities (see previous section),
in particular ``local descendants''
of globally conserved quantities like
energy, charge, particle numbers, etc.:
Examples are the amount of energy, charge etc.
within some subdomain of 
the total system, or, more generally, 
local densities, whose
content within a given volume 
can only change via transport currents 
through the boundaries of that volume.
As a consequence, the global relaxation 
process becomes ``unusually slow''
if the densities between 
macroscopically separated places
need to equilibrate (small surface-to-volume
ratio), or if there exists a 
natural ``bottleneck'' for their 
exchange 
(weakly interacting subsystems).

Put differently, our present theory 
is meant to describe the very rapid 
relaxation towards local equilibrium, 
but not any subsequent global equilibration.
Only if there exists a clear-cut time-scale 
separation between these two relaxation 
steps (or if there is no second step at all)
can we hope to quantitatively 
capture the first step by our results.
Conversely, the time scale-separation 
usually admits some Markovian 
approximation for the second step,
yielding an exponential decay,
whose time scale still depends 
on many details of the system.

Natural further generalizations 
include the closely related
concepts of hindered equilibrium,
quasi-equilibrium (metastability), 
and, above all, prethermalization
\cite{ber04,moe08,gri12},
referring, e.g., to a fast partial 
thermalization within a certain 
subset of modes, (quasi-)particles, 
or other generalized degrees of freedom.
(Like in \cite{ber04},  we do not adopt 
here the additional requirement \cite{moe08} 
that the almost conserved quantities originate 
from a weak perturbation of an integrable 
system.)

In short, our working hypothesis 
is that the theory (\ref{140})
describes the temporal relaxation of 
$\langle A\rangle_{\!\rho(t)}$
for any given pair $(\rho(0),A)$ 
unless one of them
is exceptionally close to 
or in some other way 
slowed down by an (almost) conserved 
quantity.

\subsection*{Comparison with experimental results}
We focus on experiments in closed many-body 
systems in accordance with the above 
general requirements.
In comparing them with our theory (\ref{140}),
we furthermore assume that the (pre-)thermalized
system occupies a microcanonical 
energy window
with some (effective) temperature $T$ and 
$\Delta E\gg \kB T$, so that
(\ref{210}) applies.
Finally, the 
asymptotic
values $\langle A\rangle_{\!\rho(0)}$
and $\langle A\rangle_{\!\rhomic}$
in (\ref{140}) are either obvious or
will be estimated from
the measurements, hence no further knowledge 
about the often quite involved details of the 
experimental observables will be needed! 

\begin{figure}
\epsfxsize=1.0\columnwidth
\epsfbox{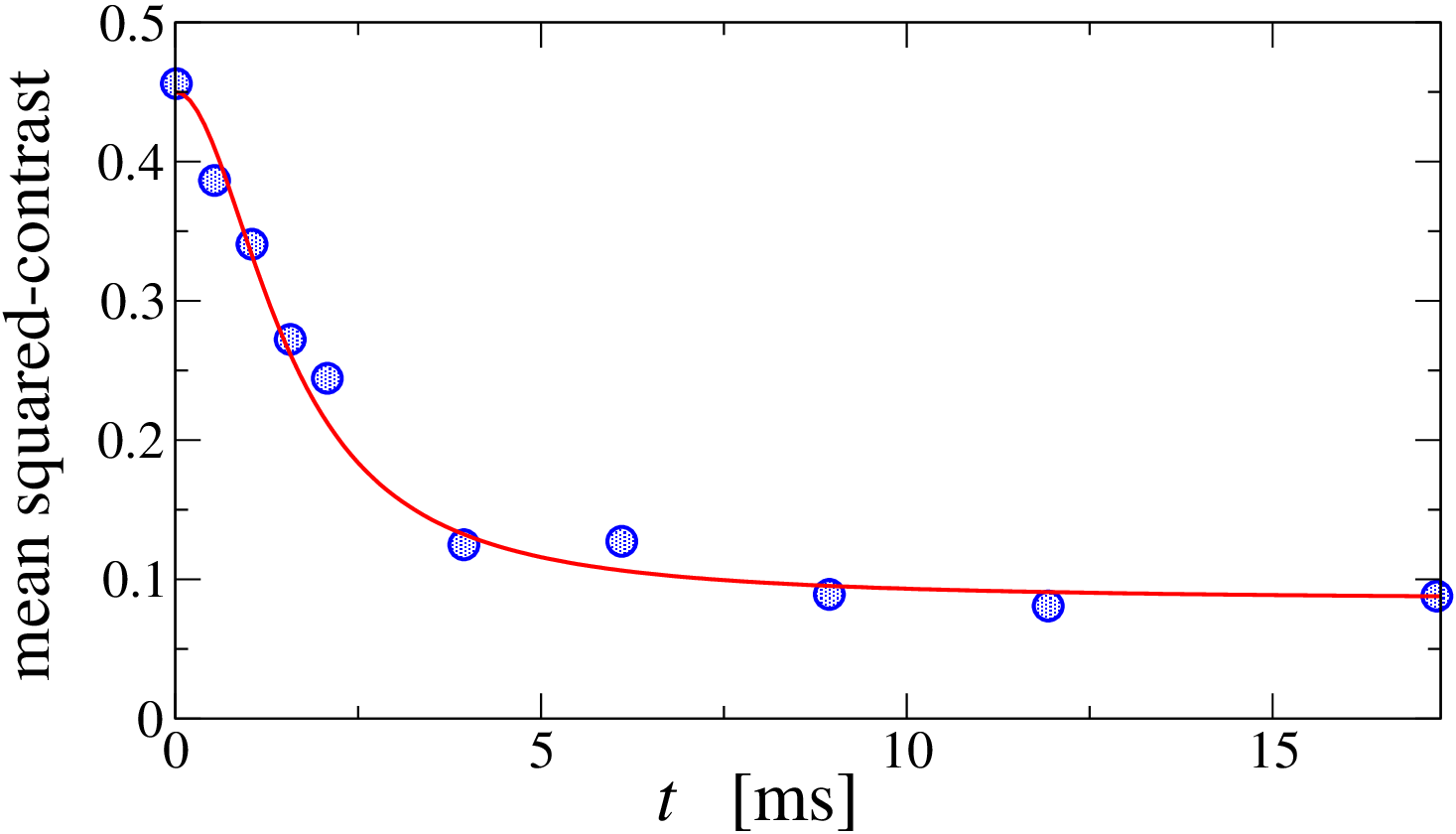}
\caption{\label{fig1}
{\bf Prethermalization of ultracold atoms.}
The considered observable ``mean squared-contrast'' 
quantifies the spatial correlation of the
matter-wave interference pattern
after coherently splitting a Bose gas into two 
quasi-condensates (see \cite{gri12} 
for more details).
Symbols: Experimental data from Fig. 2A 
of Ref. \cite{gri12}.
Line: Theoretical prediction 
(\ref{140}), (\ref{210}) 
with $T=5$\,nK.
The pertinent effective temperature
has also been roughly
estimated in Ref. \cite{gri12}
(see Fig. 2B therein)
and is still compatible with
our present fit $T=5\,$nK.
As discussed at the end of section 
``Typical fast relaxation and prethermalization'',
the depicted prethermalization is followed 
by a much slower, global thermalization
\cite{gri12}, which is omitted in the present figure.
}
\end{figure}

Fig. \ref{fig1} demonstrates the very good agreement
of the theory with the rapid initial prethermalization
of a coherently split Bose gas, observed by 
the Schmiedmayer group in Ref. \cite{gri12}.

\begin{figure}
\epsfxsize=1.0\columnwidth
\epsfbox{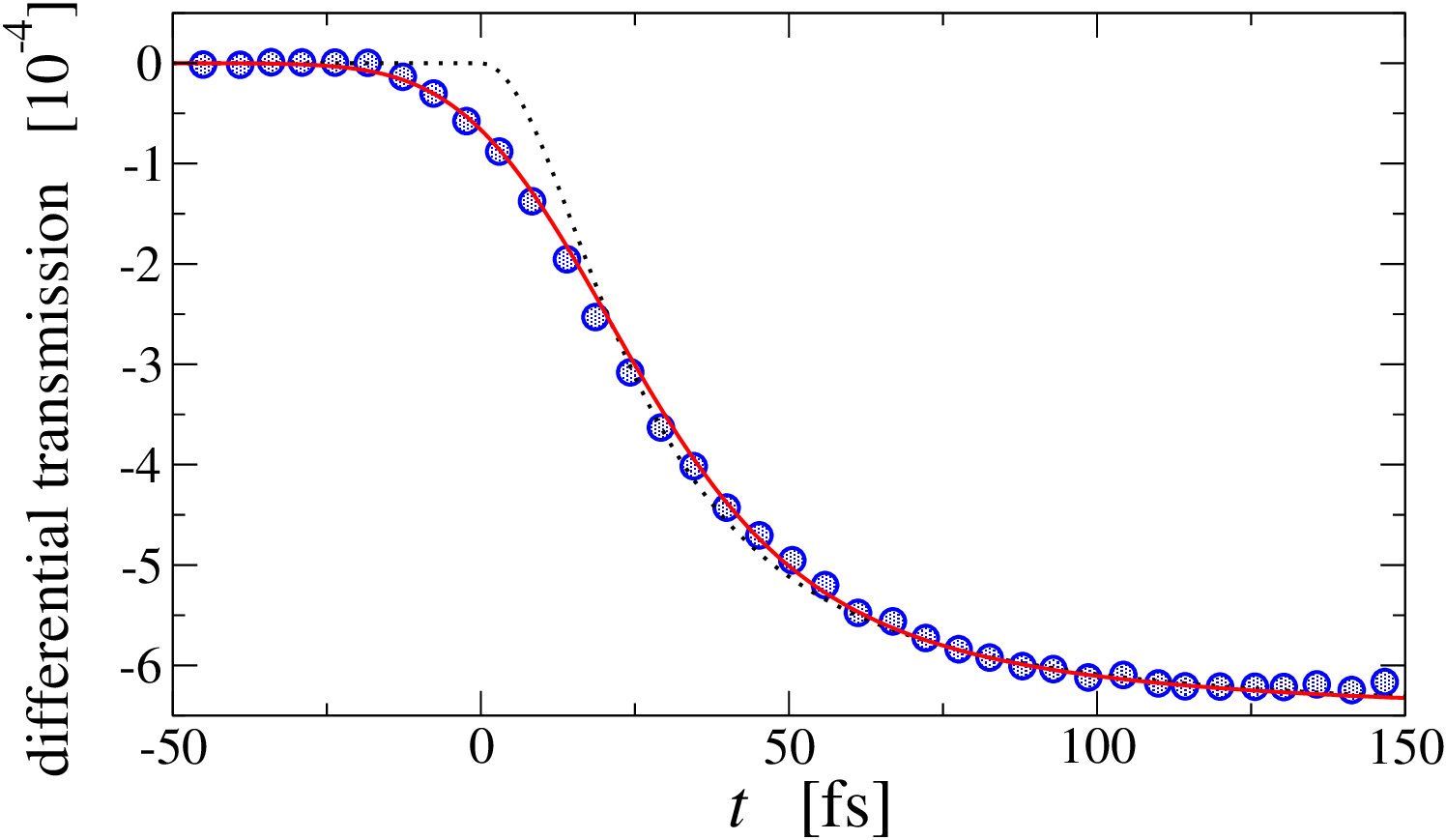}
\caption{\label{fig2}
{\bf Ultrafast relaxation of hot electrons.}
A first laser pulse (at $t=0$) ``heats up'' 
the electron gas in a thin ferromagnetic 
film, whose re-thermalization 
is then probed
by means of a second laser pulse.
As detailed in \cite{gui02}, the 
considered observable 
``differential transmission'' 
quantifies the magneto-optical 
polarization rotation of the 
probe laser light.
Symbols: Experimental data from Fig. 2a
of Ref. \cite{gui02}.
Dotted: Theoretical prediction 
(\ref{140}), (\ref{210}) 
with $T=310$\,K and $F(t<0):=1$.
Solid: Convolution of the dotted line 
with a Gaussian of $35$\,fs FWHM,
accounting for the finite widths of the 
pump and probe laser pulses (see also main text).
Similarly as in Fig. \ref{fig1}, 
on larger time-scales than covered
by the present figure, the prethermalized 
electrons also exhibit non-negligible 
interactions 
with the lattice phonons and magnons,
resulting in a much slower global relaxation 
of the compound electron-lattice 
system \cite{gui02}.
Concerning the pertinent temperature 
$T$, a direct experimental estimate
is not available for the setup from 
Ref. \cite{gui02}
(I contacted one of the authors),
but it has been provided
for a similar experiment by the same 
group in Ref. \cite{beau96},
except that the fluence 
(energy per spot area of the pump laser pulse) 
was 70 times larger than in \cite{gui02}.
Taking all this into account,
the estimate $T=310$\,K adopted in 
the present figure 
seems very reasonable.
}
\end{figure}

In Fig. \ref{fig2}, the theory is compared with
the pump-probe experiment by the 
Bigot group from Ref. \cite{gui02}. 
The finite widths of the pump and the 
probe laser pulses are
roughly accounted for by convoluting  
equation (\ref{140}) with a Gaussian of 
$35$\,fs FWHM (Full Width at Half Maximum).
In Ref. \cite{gui02}, the FWHM 
of the pump pulse is estimated as $20$\,fs
and the combined FWHM for both pulses as $22$\,fs,
implying a FWHM of $9$\,fs for the probe pulse.
The latter value seem quite optimistic to us.
A second ``excuse'' for our slightly 
larger FWHM value of $35$\,fs 
is that the tails of the experimental 
pulse shape may be considerably broader than
those of a Gaussian with the same FWHM 
(see, e.g., Fig. 2c in the supplemental 
material of Ref. \cite{gie15}).
Finally, the convolution of (\ref{140}) 
with a Gaussian represents a rather 
poor ``effective description'' in the 
first place:
Our entire theoretical approach 
becomes strictly speaking invalid when
the duration of the perturbation becomes
comparable to the thermalization time.

\begin{figure}
\epsfxsize=0.9\columnwidth
\epsfbox{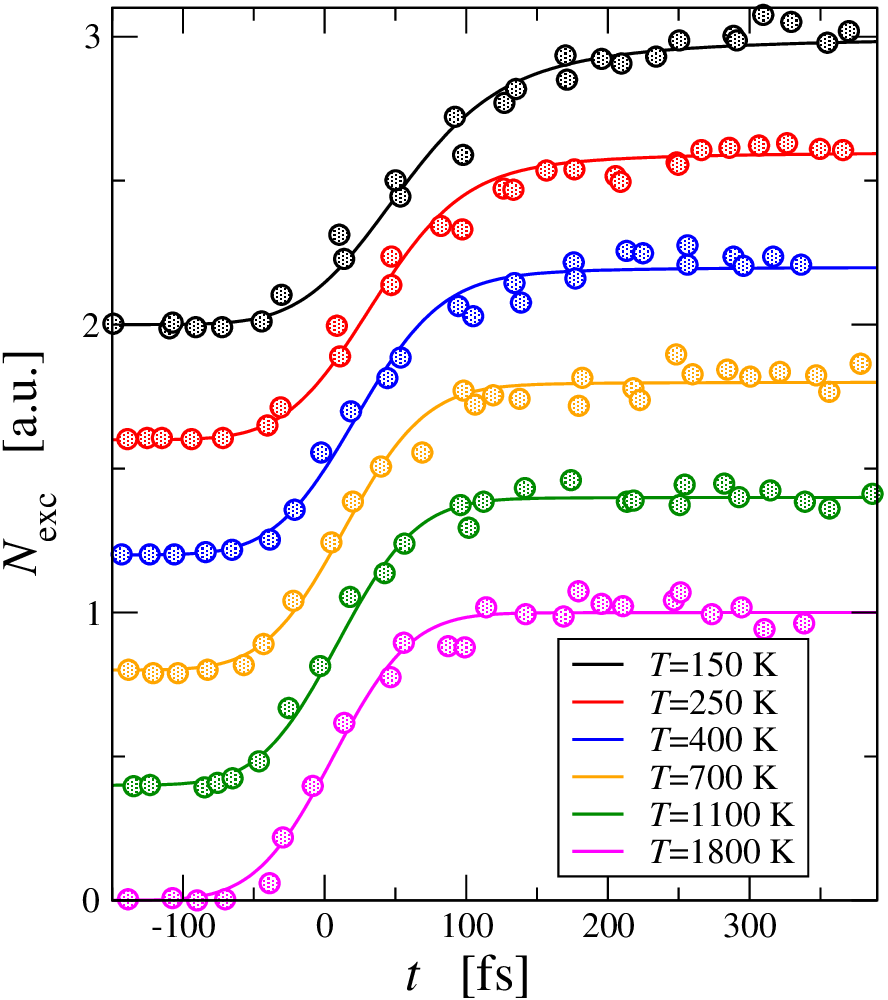}
\caption{\label{fig3}
{\bf Temperature dependent relaxation of hot electrons.}
Symbols: Similar pump-probe experiments 
as in Fig. \ref{fig2}, but now conducted 
on bismuth and for 6 different fluences
(energy per spot area of the pump laser pulses). 
As detailed in \cite{fau13}, the considered
observable $N_{\rm{exc}}$ 
quantifies (in arbitrary units) the number of 
excited electrons above the Fermi level.
The depicted data are from Fig. 5b of Ref. \cite{fau13}
for fluences (top-down)
0.12, 0.2, 0.36, 0.52, 0.68, and 0.84 mJ\,cm$^{-2}$.
Lines: Theoretical prediction (\ref{140}), (\ref{210})
with temperatures as indicated and
convoluted with a Gaussian 
of $100$\,fs FWHM (see also main text).
The conversion of a given fluence into a
temperature change of the electron gas is not obvious.
In particular, the estimates provided in \cite{fau13}
seem not very reliable to us: 
First of all, Fig. 6 in \cite{fau13}
indicates a temperature of ca. $250$\,K at 4
different time-points about $200$\,fs before 
the pump pulse, while the actual temperature
of the unperturbed system is known to be $130$\,K.
Second, the temperature error bars in 
Fig. 6b of \cite{fau13} are quite large.
Third, a key premise of those estimates 
in \cite{fau13} is that the ``renormalized'' 
curves in Fig. 3S(B) of \cite{pap12} should coincide,
while their actual agreement is only moderately
better than for the ``bare'' curves in Fig. 3S(A).
For all these reasons, we used the temperature 
as a fit parameter in 
the present figure.
}
\end{figure}

A similar comparison 
with the pump-probe experiments 
by Faure at al. from Ref. \cite{fau13} is
presented in Fig. \ref{fig3}.
As before, we adopted a slightly larger FWHM of
$100$\,fs than the estimate of $76$\,fs in \cite{fau13}.
Due to the above mentioned fundamental limitations
of our theory for such rather large FWHM 
values, the temperatures adopted in Fig. \ref{fig3}
should still be considered as quite crude estimates.
Apart from that, Fig. \ref{fig3} nicely confirms
the predicted temperature dependence from
(\ref{210}).

We close with three remarks:
First, Refs. \cite{gui02,fau13} also 
implicitly confirm our prediction
that the essential temporal relaxation
(encapsulated by $F(t)$ in (\ref{140}))
is generically the same for different 
observables.
Second, similar pump-probe experiments
abound in the literature, but usually
the pulse-widths are too large for our
purposes.
Third, the temporal relaxation in Figs. 
\ref{fig1}-\ref{fig3}  has also been 
investigated numerically,
but closed analytical results 
have not been available before 
\cite{gri12,fau13}.

\subsection*{Comparison with numerical results}
\begin{figure}
\epsfxsize=0.92\columnwidth
\epsfbox{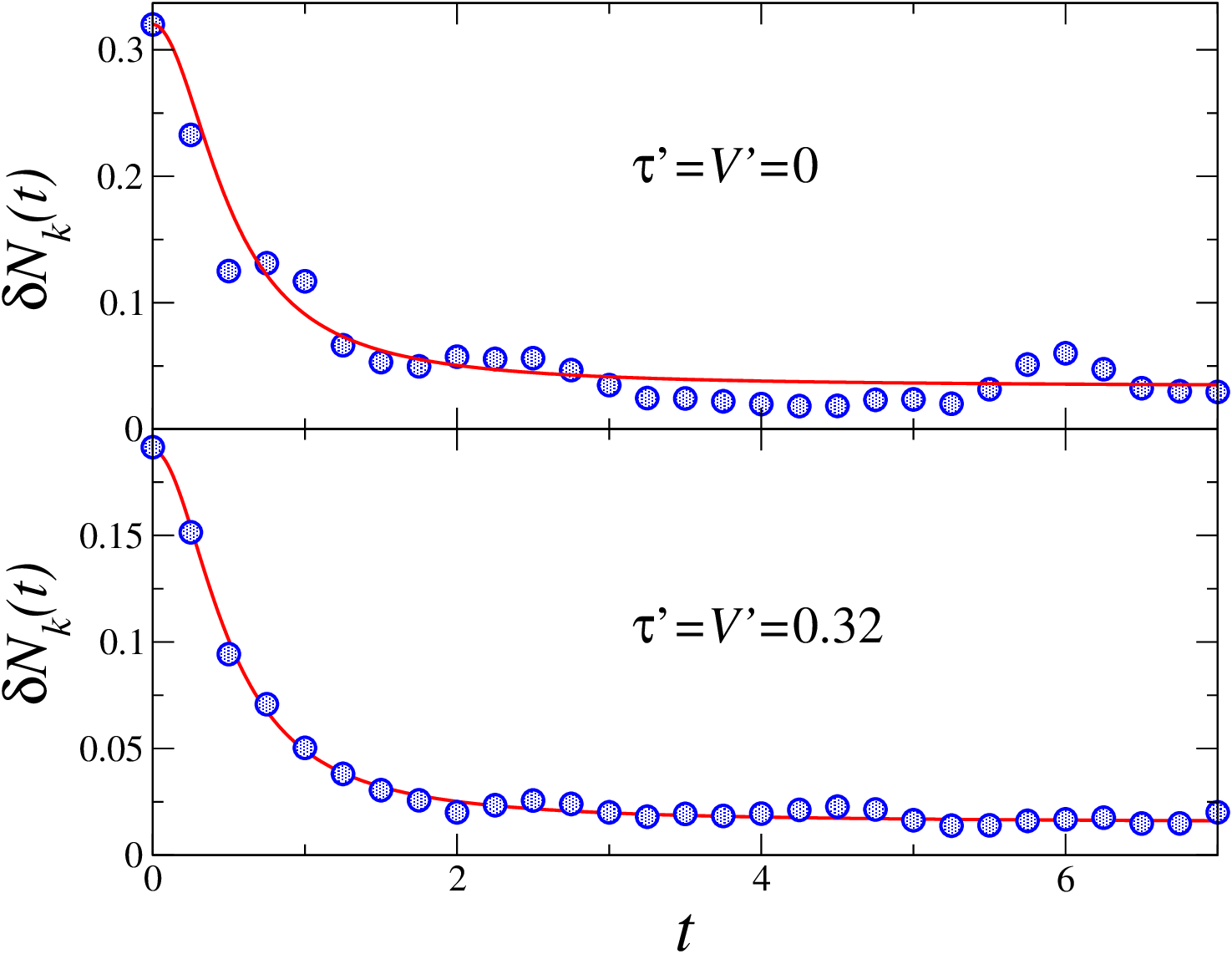}
\caption{\label{fig4}
{\bf Relaxation of an integrable (top) and
a non-integrable (bottom) fermionic model.}
Symbols: Numerical results from Ref. \cite{rig09a} 
for 8 strongly correlated fermions on 
a one-dimensional lattice with 24 sites, 
described in terms of an extended Hubbard model with 
nearest- and next-nearest-neighbor hopping and 
interaction parameters $\tau$, $\tau'$, $V$, 
and $V'$, respectively.
Working in units with $\hbar=\kB =\tau=V=1$
and focusing on parameters $\tau'=V'$, 
the model is integrable if $\tau'=V'=0$
and non-integrable 
otherwise.
A quantum quench generates
an initial pure state out of equilibrium,
whose energy corresponds
to that of a canonical ensemble 
with temperature $T=2$.
As detailed in \cite{rig09a}, 
the considered observable 
$\delta N_k(t)$ is a dimensionless
descendant of the density-density 
structure factor.
The depicted data are from Fig. 1(g),(j) 
of Ref. \cite{rig09a}.
Lines: Theoretical predictions (\ref{140}), (\ref{210})
with $T=2$. 
}
\end{figure}

Fig. \ref{fig4} illustrates the very good 
agreement of our theory 
with Rigol's numerical findings
from Ref. \cite{rig09a},
both for an integrable 
and an non-integrable example.
A similar agreement is found for
all other parameters and also for 
an analogous hardcore boson model 
examined in Refs. \cite{rig09b,rig09a}.
On the other hand, a second observable
considered in Ref. \cite{rig09a}, deriving
from the momentum distribution function,
exhibits in all cases a significantly
slower and also qualitatively different
temporal relaxation.
According to the discussion in section
``Typical fast relaxation and prethermalization'',
it is quite plausible that the latter
observable is indeed ``non-typical'' 
in view of the fact that it represents
a conserved quantity for fermions 
with $V=\tau'=V'=0$ \cite{rig09a}.

\begin{figure}
\epsfxsize=0.92\columnwidth
\epsfbox{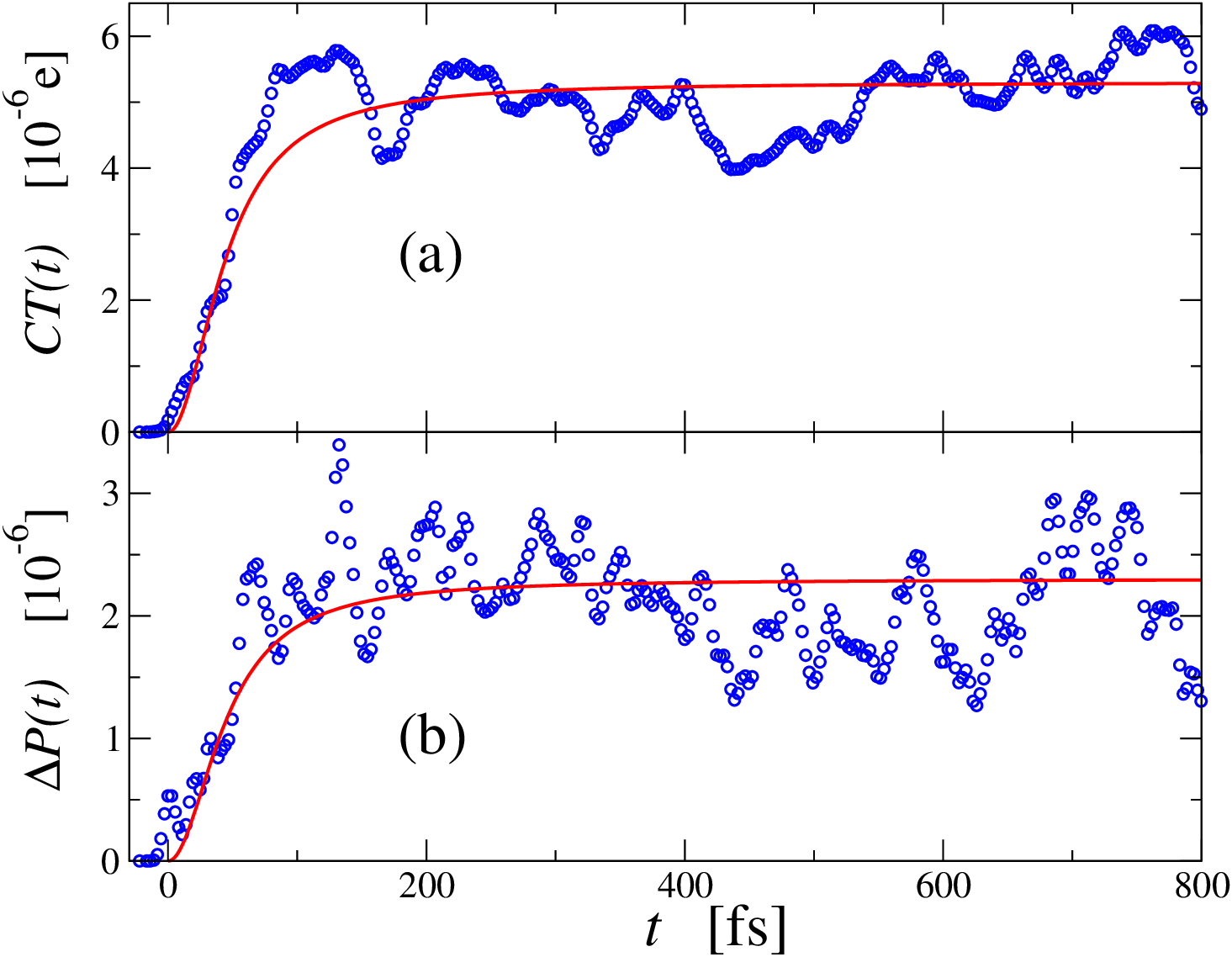}
\caption{\label{fig5}
{\bf Prethermalization in a one-dimensional electron gas.}
Symbols: Numerical results from Ref. \cite{tho04} for a 
one-dimensional model of the 
many-electron dynamics in an asymmetric 
double-well potential 
(emulating a metal-insulator-metal junction).
Starting with 44 electrons in the ground 
state, a laser pulse-like electrical perturbation 
acts predominantly 
on the 16 electrons in the smaller,
box-shaped well, 
and then their re-thermalization is 
monitored via the charge transfer into the 
larger well (denoted in (a) as $CT(t)$),
and via the change of the ground 
state population (denoted in (b) as $\Delta P(t)$).
Depicted are the numerical results from Fig. 8 
of Ref. \cite{tho04}.
For further details regarding the simulations 
we refer to \cite{tho04,kla03}. 
Lines: Theoretical predictions (\ref{140}), 
(\ref{210}), exploiting the estimate 
$T=170$\,K from the main text, 
and neglecting the finite temporal width 
($20$\,fs) of the pulse.
As in Figs. \ref{fig1}-\ref{fig3}, 
we are actually dealing with a prethermalization 
process within the smaller well. 
The subsequent global thermalization is much 
slower due to the high barrier between the wells.
Considering that $\langle A\rangle_{\!\rhomic}$ 
is the only remaining fit parameter 
in the theory from (\ref{140}) and (\ref{210}), 
the agreement with the simulations
is remarkably good.
In particular, the two very different observables 
$CT(t)$ and $\Delta P(t)$ are indeed governed 
by the same $F(t)$, as predicted 
by (\ref{140}), (\ref{210}).
}
\end{figure}

In Fig. \ref{fig5} we compare
our theory with the simulations of a different 
one-dimensional electron model by Thon et al. 
from Ref. \cite{tho04}.
In doing so, the pertinent temperature $T$ has
been estimated as follows:
The textbook Sommerfeld-expansion for 
$N$ electrons in a one-dimensional box 
yields $E=E_0[1+(3\pi^2/8)(\kB T/\EF)^2]$,
where $E$ is their total energy, $E_0=(1/3)N\EF$ 
the ground state energy, $\EF=(\pi\hbar N/gL)^2/2m$
the Fermi-energy, $L$ the box length, $m$ the
electron mass, and $g:=2s+1=2$ ($s=1/2$ for electrons).
Assuming that the pulse acts solely on the small well
implies $N=16$, $L\simeq 15$\,nm \cite{tho04},
and $E-E_0\simeq 0.045$\,eV (see Fig. 8a in \cite{tho04}).
Altogether, we thus obtain $T\simeq 170$\,K.

The remnant ``fluctuations'' of the numerical data
in Figs. \ref{fig4} and \ref{fig5} can be readily 
explained as finite particle number effects
(see Fig. 4 in \cite{rig09a} 
and Fig. 10 in \cite{tho04}),
and their temporal correlations are as 
predicted below equation (\ref{140}).
The seemingly rather strong fluctuations 
in Fig. \ref{fig5} are a fallacy since the 
systematic changes themselves are very small.

\begin{figure}
\epsfxsize=1.0\columnwidth
\epsfbox{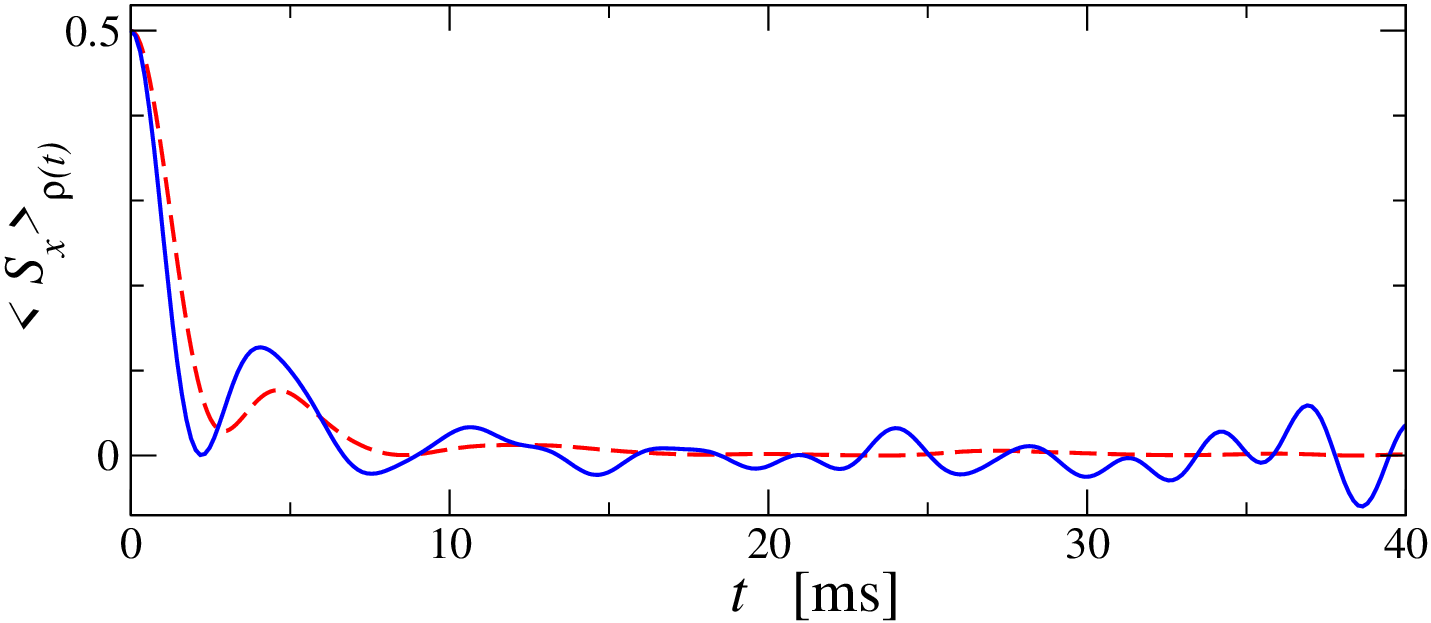}
\caption{\label{fig6}
{\bf Thermalization of a spin qubit coupled to a bath.}
Solid: Numerical results for the
model with 7 spin-1/2 degrees 
of freedom in an external magnetic 
field from Ref. \cite{het15}:
A central spin (qubit) is randomly 
(and reasonably weakly) 
coupled to a bath of 6 spins.
The initial state $\rho(0)$ is the 
product of a totally mixed bath state 
and an eigenstate of the central spin
component $S_x$.
Depicted are the data from Fig. 2 of 
Ref. \cite{het15} for the central spin 
component $S_x$.
Dashed: Theoretical prediction 
(\ref{140}), (\ref{150}), (\ref{90}).
Due to the above mentioned initial 
condition and the quite small dimension $D=2^7$, 
the approximations (\ref{190})-(\ref{210}) 
are not very well satisfied by the actual 
energy eigenvalues $E_1$,...,$E_{128}$
(kindly provided by the authors of 
Ref. \cite{het15}).
Hence, we have evaluated $F(t)$ in 
(\ref{140}) directly via (\ref{150})
and (\ref{90}).
}
\end{figure}

Next we turn to the numerical findings 
for a qubit in contact with a spin bath
by the Trauzettel group from Ref. \cite{het15}.
The agreement with our theory in Fig. \ref{fig6} 
is as good as it possibly can be for such a 
rather small dimensionality of $D=2^7$.
Indeed, the remaining differences nicely confirm
the predictions below equation (\ref{140}),
regarding both their typical order of magnitude 
$\da\sqrt{\tr\{\rho^2(0)\}/\DD}
=1\sqrt{2^{-6}/2^7} \simeq 0.01$
and their temporal correlations
(where we exploited that $\tr\{\rho^2(0)\}=2^{-6}$
for the particular initial condition 
$\rho(0)$ adopted in Fig. \ref{fig6}).

\begin{figure}
\epsfxsize=1.0\columnwidth
\epsfbox{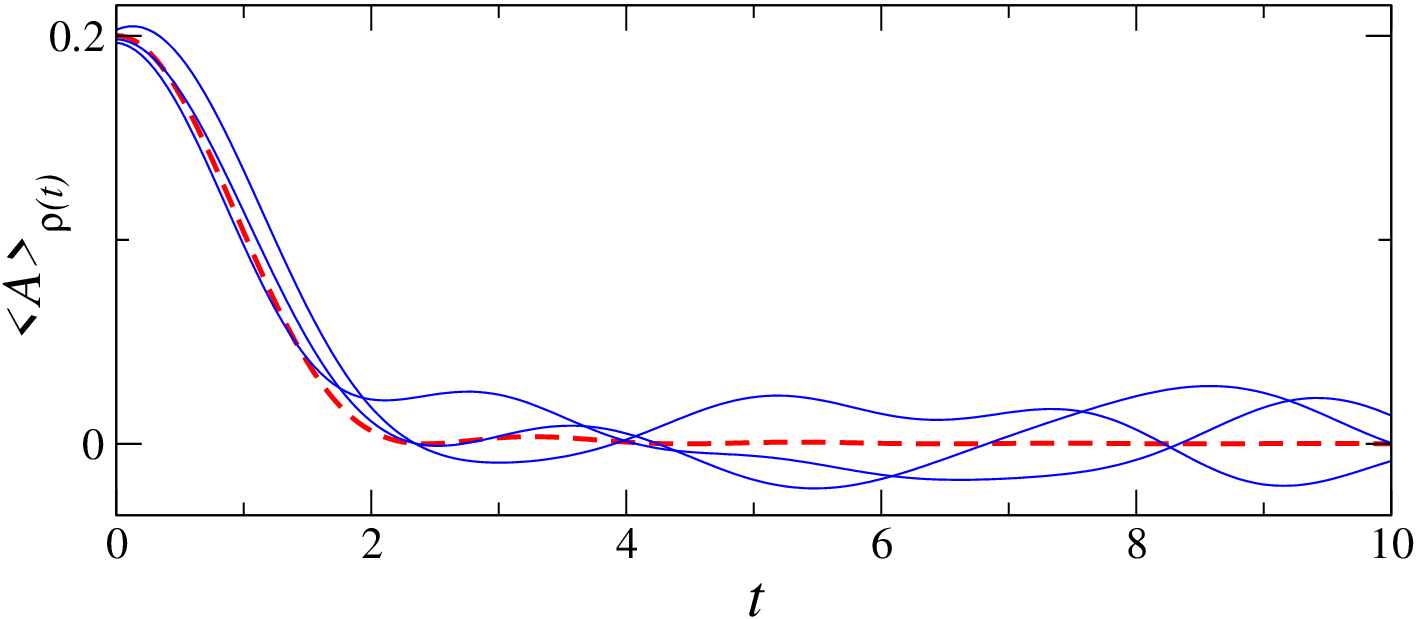}
\caption{\label{fig7}
{\bf Thermalization in a random matrix model.}
Solid: Numerical results for the random matrix 
model of the form $H=H_0+\lambda V$ from 
Ref. \cite{bar09}.
Adopting dimensionless units with $\hbar=1$,
the $\DD=6000$ eigenvalues of $H_0$ are 
chosen equidistant with level spacing 
$8.33\cdot 10^{-5}$ \cite{bar09}.
The matrix elements of $A$ (observable)
and $V$ (perturbation) in the basis 
of $H_0$ satisfy $A_{ik}=(-1)^k\delta_{ik}$
and $V_{ki}=V_{ik}^\ast$.
Apart from the latter constraint,
the real and imaginary parts of $V_{ik}$
are independent, normally distributed 
random numbers.
The initial state is 
$\rho(0)=|\psi\rangle\langle\psi|$,
where $|\psi\rangle$ is randomly sampled 
from the energy shell $\hr$
under the constraint 
$\langle A\rangle_{\!\rho(0)}\simeq 0.2$ \cite{bar09}.
Depicted are three representative numerical 
realizations for $\lambda = 7\cdot 10^{-3}$
akin to Fig. 1b of Ref. \cite{bar09}
(in dimesionless units).
Dashed: Theoretical prediction (\ref{140}), (\ref{150}), (\ref{90}).
Similarly as in Fig. \ref{fig6}, the numerically 
obtained energies $E_1,...,E_{6000}$ were found to 
satisfy (\ref{190})-(\ref{210}) not very well, 
hence we have directly evaluated (\ref{90}), (\ref{150}).
}
\end{figure}

Our final example is Bartsch and Gemmer's 
random matrix model from Ref. \cite{bar09}.
Referring to the notation and definitions
in the caption of Fig. \ref{fig7}, 
one readily sees that the considered 
observable $A$ is a conserved quantity for 
the unperturbed Hamiltonian ($\lambda =0$).
In agreement with our discussion in section
``Typical fast relaxation and prethermalization'',
$A$ is therefore still ``almost conserved''
for small $\lambda$ and indeed exhibits 
a slow, exponential decay towards 
$\langle A\rangle_{\!\rhomic}=0$
(see Fig. 1a in \cite{bar09}).
Upon increasing $\lambda$, one recovers the 
much faster, non-exponential decay of our 
present theory (see Fig. 1b in \cite{bar09}).
Unfortunately, the $\lambda$-value $1.77\cdot 10^{-3}$ 
from Fig. 1b of \cite{bar09} is still somewhat 
too small and the eigenvalues $E_1,...,E_{6000}$ 
are not any more available (I asked the authors).
Therefore, we repeated the numerics from 
\cite{bar09} on our own for $\lambda = 7\cdot 10^{-3}$.
The resulting agreement with (\ref{140}) in Fig. \ref{fig7} 
is very good, and the temporal correlations of the 
deviations as well as their typical order of magnitude
$\da\sqrt{\tr\{\rho^2(0)\}/\DD}=2\sqrt{1/6000}\simeq 0.03$
are as predicted below (\ref{140}).

We close with two remarks:
First, there is no fit parameter 
in any of the above examples apart from
$\langle A\rangle_{\!\rho(0)}$ in Fig. \ref{fig4}
and $\langle A\rangle_{\!\rhomic}$ in Figs.
\ref{fig4} and \ref{fig5}.
Second, especially in the case of the integrable model
in Fig. \ref{fig4}, one may question whether the
considered system exhibits thermalization in the
first place, as is tacitly assumed in equation 
(\ref{140}). In Supplementary Note 2 we argue 
that  (\ref{140}) indeed is expected to still remain 
valid in such cases if $\langle A\rangle_{\!\rhomic}$ 
is replaced by the pertinent non-thermal
long-time asymptotics (which, in turn, 
is estimated from the numerical data 
in Fig. \ref{fig4}).

\section*{Discussion}
Our main result (\ref{140}) implies thermalization 
in the sense that a generic non-equilibrium system
with a macroscopically well defined energy 
becomes practically indistinguishable from
the corresponding microcanonical 
ensemble for the overwhelming majority of 
all sufficiently late times.
Apart from the concrete initial and long-time expectation values 
(i.e. $\langle A\rangle_{\!\rho(0)}$ and 
$\langle A\rangle_{\rhomic}$ 
in (\ref{140})), the temporal relaxation
(i.e. $F(t)$ in (\ref{140})) 
depends only on the
spectrum of the Hamiltonian within the pertinent 
interval of non-negligibly populated 
energy eigenstates, but 
not on any further details of the initial condition 
or the observable.
This represents one of the rare instances of a 
general quantitative statement about systems 
far from equilibrium.

The theory agrees very well with a wide variety of 
experimental and numerical results
from the literature
(though none of them was originally conceived
for the purpose of such a comparison).
We are in fact not aware of any other
quantitative analytical explanation of 
those data comparable to ours.
Indeed, the usual paradigm to identify and 
then analytically quantify the main physical 
mechanisms seems almost hopeless here.
In a sense, our present approach thus 
amounts to a new paradigm:
There is no need of any further ``explanations''
since the observed behavior is expected 
with overwhelming likelihood from the very 
beginning, i.e., unless there are special 
{\em a priori} reasons to the contrary.

Similarly as in \cite{mal14,gol15,gol15b,cra12},
generic thermalization is found to happen extremely
quickly (unless the system's energy or temperature
is exceedingly low). 
Moreover, the temporal decay is 
typically non-exponential.
A main prediction of our theory is that these features
should in fact be very common
(at least in the form of prethermalization), 
but often they are unmeasurably fast 
or they have simply not been looked for
so far.
Conversely, most of the usually
considered observables and initial 
conditions are actually quite
``special'', namely exceptionally slow, 
``almost conserved'' quantities.
A better understanding of those principally
untypical but practically very common 
thermalization processes remains an 
open problem \cite{mal14,gol15,gol15b}.

\section*{METHODS} 
\subsection*{Basic matrices}
According to section ``Analytical results'',
the unitary $U$ represents the basis 
transformation between the eigenvectors 
$|n\rangle$ ($n=1,...,\DD$)
of the Hamiltonian $H$ and those 
of the observable $A$.
Denoting the eigenvalues of $A$ by $\lambda_\nu$ 
and the eigenvectors by $|\psi_\nu\rangle$ ($\nu=1,...,\DD$), 
the matrix elements of $U$ are thus
$U_{n\nu}:=\langle n|\psi_\nu\rangle$.
Accordingly, the matrix elements of 
$\rho(0)$ in the basis of $H$ are 
related to those in the basis of 
$A$ via
\begin{eqnarray}
\rho_{mn}(0)
=
\sum_{\mu,\nu=1}^{\DD}
U_{m\mu} \, \rho_{\mu\nu}\, U_{n\nu}^\ast \ ,
\label{5b}
\end{eqnarray}
where $\rho_{\mu\nu}:=\langle \psi_\mu|\rho(0)|\psi_\nu\rangle$.
Similarly, the matrix elements of $A$ satisfy
\begin{eqnarray}
A_{mn}
=
\sum_{\xi=1}^{\DD}
U_{m\xi}\, \lambda_\xi \,  U_{n\xi}^\ast
\label{6b}
\end{eqnarray}
and hence
\begin{eqnarray}
\rho_{mn}(0)A_{nm}
=
\sum_{\mu,\nu,\xi=1}^{\DD}
\rho_{\mu\nu}\, \lambda_\xi \,
U_{m\mu} U_{n\nu}^\ast U_{n\xi} U_{m\xi}^\ast \ .
\label{10b}
\end{eqnarray}

As announced below equation (\ref{30}), we
work (without loss of generality)
in a reference frame (or reference basis of $\hr$)
so that only $H$ (and thus $|n\rangle$) depends on
$U$, while $A$ and $\rho(0)$ 
(and thus $|\psi_\nu\rangle$)
are independent of $U$.
Hence, $\rho_{\mu\nu}$ and 
$\lambda_\xi$ on the right hand side of
equations (\ref{5b})-(\ref{10b}) are 
independent of $U$.

\subsection*{Derivation of equation (\ref{100})}
As a simple first exercise, let us average equation (\ref{10b}) 
over all uniformly (Haar) distributed unitaries $U$,
as specified in 
section ``Analytical results''.
Since the factors $\rho_{\mu\nu}\lambda_\xi$ on the 
right hand side are independent of $U$, we are 
left with averages over the $U$ matrix elements.
Such averages have been evaluated repeatedly 
and often independently of each other in the literature, 
see e.g. \cite{bro81,col06,gem09,bro96},
a key ingredient being symmetry arguments due to
the invariance of the Haar measure under arbitrary 
unitary transformations.
Particularly convenient for our present purposes
is the formalism adopted by Brouwer and Beenakker,
see Ref. \cite{bro96} and further references therein.
The general structure of such averages is provided
by equation (2.2) in \cite{bro96}, reading
\begin{eqnarray}
& & \lu U_{a_1b_1} \ldots U_{a_mb_m} U^\ast_{\alpha_1\beta_1} \ldots 
U^\ast_{\alpha_n \beta_n} \ru
=
\nonumber
\\
& & 
=\delta_{mn}
\sum\limits_{P,P'}V_{P,P'} \prod_{j=1}^n 
\delta_{a_j\alpha_{P(j)}} \delta_{b_j\beta_{P'(j)}} \ .
\label{20b}
\end{eqnarray}
Quoting verbatim from Ref. \cite{bro96}, 
``the summation is over all permutations 
$P$ and $P'$ of the numbers $1,...,n$.
The coefficients $V_{P,P'}$ depend only on the cycle structure 
of the permutation $P^{-1}P'$. Recall that each permutation of 
$1,...,n$ has a unique factorization in disjoint cyclic 
permutations (``cycles'') of lengths $c_1,....,c_k$ 
(where $n=\sum_{j=1}^k c_j$).
The statement that $V_{P,P'}$ depends only on the cycle structure of
$P^{-1}P'$ means that $V_{P,P'}$ depends only on the lengths 
$c_1,...,c_k$ of the cycles
in the factorization of $P^{-1}P'$.
One may therefore write $V_{c_1,...,c_k}$ instead of $V_{P,P'}$.''
The explicit numerical values of all $V_{c_1,...,c_k}$ 
with $n \leq 5$ are provided by the columns ``CUE'' of 
Tables II and IV  in \cite{bro96}.
Further remarks: The labels $m$ and $n$ in 
(\ref{20b}) have nothing to do with those in (\ref{10b}).
Equation (\ref{20b}) equals zero unless $m=n$.
Every label $a_j$ must have a ``partner'',
i.e., its value must coincide with 
one of the $\alpha_j$'s,
and vice versa, since otherwise the product 
over the Kronecker delta's $\delta_{a_j\alpha_{P(j)}}$
in (\ref{20b}) would be zero for all $P$'s.
Note that some $a_j$'s may assume the same value, 
but then an equal number of $\alpha_j$'s also must 
assume that value.
Likewise, every $b_j$ needs a
``partner'' among the $\beta_j$'s, 
and vice versa.

Adopting the abbreviation
\begin{eqnarray}
X_{mn}:=\lu \rho_{mn}(0)A_{nm} \ru
\label{30b}
\end{eqnarray}
and the renamings $a_1:=m$, $a_2:=n$, $b_1:=\mu$, $b_2:=\xi$, $b_3:=\nu$,
equation (\ref{10b}) yields
\begin{eqnarray}
X_{a_1a_2}=\sum_{b_1,b_2,b_3} \rho_{b_1 b_3} \lambda_{b_2} \lu 
U_{a_1 b_1} U_{a_2 b_2} U_{a_1 b_2}^\ast U_{a_2 b_3}^\ast 
\ru \ .
\label{40b}
\end{eqnarray}
The connection with (\ref{20b}) is established via
the identifications $\alpha_1:=a_1$, $\alpha_2:=a_2$,
$\beta_1:=b_2$, $\beta_2:=b_3$.
Therefore, if $b_1\not=b_2$ then 
the only potential ``partner'' 
of $b_1$ is $\beta_2$, and only if their values
coincide, i.e. $b_3=b_1$, the corresponding 
summands may be non-zero.
The same conclusion can be drawn if $b_1=b_2$.
We thus can rewrite (\ref{40b}) with (\ref{20b}) as
\begin{eqnarray}
X_{a_1a_2}=\sum_{b_1,b_2} \rho_{b_1 b_1} \lambda_{b_2} \sum\limits_{P,P'}
V_{P,P'} \prod_{j=1}^2 \delta_{a_j a_{P(j)}} \delta_{b_j\beta_{P'(j)}} 
\label{50b}
\end{eqnarray}
where $\beta_1=b_2$ and $\beta_2=b_1$.

There are two permutations of the numbers $1,2$,
namely the identity and one, which exchanges $1$ 
and $2$.
Denoting them as $P_1$ and $P_2$, respectively,
and observing that $\beta_j=b_{P_2(j)}$, 
equation (\ref{50b}) can be rewritten as
\begin{eqnarray}
X_{a_1a_2} & = & \sum\limits_{k=1}^2 \prod_{j=1}^2 \delta_{a_j a_{P_k(j)}}
\sum\limits_{l=1}^2 V_{P_k,P_l} S_l
\label{60b}
\\
S_l & := & \sum_{b_1,b_2} \rho_{b_1 b_1} \lambda_{b_2}  
\prod_{j=1}^2 \delta_{b_j b_{P_2(P_l(j))}} 
\label{70b}
\end{eqnarray}
For $l=1$ the two Kronecker delta's in (\ref{70b})
both require that $b_1=b_2$ and hence
\begin{eqnarray}
S_1 =\sum_{b_1} \rho_{b_1b_1} \lambda_{b_1} = \tr\{\rho(0)A\} \ .
\label{80b}
\end{eqnarray}
The last equality can be verified by 
evaluating the trace 
in the eigenbasis of $A$, see above 
equation (\ref{5b}). 
In the same way, one finds that
\begin{eqnarray}
\!\!
S_2 
= \sum_{b_1,b_2} \rho_{b_1 b_1} \lambda_{b_2}
= \tr\{\rho(0)\} \tr\{A\} = \DD \tr\{\rhomic A\} \, .
\label{90b}
\end{eqnarray}
In the last equation, we exploited that $\tr\{\rho(0)\}=1$ and
$\rhomic:=I/\DD$, see below equation (\ref{100}).
Observing that the  two Kronecker delta's in (\ref{60b})
equal one if $k=1$ or if $k=2$ and $a_1=a_2$, 
the overall result is
\begin{eqnarray}
X_{a_1a_2} & = &  
\langle A\rangle_{\! \rho(0)} ( V_{P_1,P_1}+\delta_{a_1 a_2}  V_{P_2,P_1})
\nonumber
\\
& + & 
\DD \langle A\rangle_{\! \rhomic} ( V_{P_1,P_2}+\delta_{a_1 a_2} V_{P_2,P_2} )
\  ,
\label{100b}
\end{eqnarray}
where, as usual, 
$\langle A\rangle_{\!\rho(0)}:=\tr \{\rho(0)A\}$ 
and
$\langle A\rangle_{\! \rhomic}:=\tr\{\rhomic A\}$.

Finally, the coefficients $V_{P_k,P_l}$ are evaluated as 
explained below equation (\ref{20b}):
If $k=l$ then $P_l^{-1}P_k=P_1$ factorizes 
in two cycles of lengths $c_1=c_2=1$, 
i.e. $V_{P_k,P_l}=V_{c_1,c_2}=V_{1,1}$.
Likewise, if $k\not =l$ then $P_l^{-1}P_k=P_2$ consists
of one cycle with $c_1=2$, i.e. $V_{P_k,P_l}=V_{2}$.
Referring to columns ``CUE'' and rows ``$n=2$'' 
of Tables II and IV in Ref. \cite{bro96} yields
$V_{1,1}=1/(\DD^2-1)$ and $V_2=-1/[\DD(\DD^2-1)]$.
Returning to the original labels $m$ and $n$ in 
equation (\ref{30b}), we thus can
rewrite (\ref{100b}) as
\begin{eqnarray}
\!\!
X_{mn} & = &  
\langle A\rangle_{\! \rho(0)}\frac{\DD - \delta_{mn}}{\DD(\DD^2-1)}
+
\langle A\rangle_{\! \rhomic}\frac{\DD \delta_{mn} -1}{\DD^2 - 1}
\, .
\label{110b}
\end{eqnarray}

As a consequence, we can infer from equations
(\ref{50}) and (\ref{30b}) that 
$\langle A\rangle_{\!\rhou}=\DD X_{nn}$ and 
with (\ref{110b}) that
\begin{eqnarray}
\!\!
\langle A\rangle_{\!\rhou} & = &  
\langle A\rangle_{\! \rho(0)}\frac{1}{\DD + 1}
+
\langle A\rangle_{\! \rhomic}\frac{\DD}{\DD + 1}
\  .
\label{120b}
\end{eqnarray}
Hence, one readily recovers equation (\ref{100}).

A relation remarkably similar to our 
present equation (\ref{100}), albeit in a 
quite different physical context, has been 
previously obtained also in Ref. \cite{ols12} 
(see equation (2) therein).

\subsection*{Derivation of equation (\ref{120})}
Without any doubt, there are much faster ways 
to obtain equations (\ref{110b}) or (\ref{120b}). 
The advantage of our present way is that it 
can be readily adopted without any conceptual 
differences (albeit the actual calculations 
become more lengthy) to more demanding
cases like
\begin{eqnarray}
\lu \xi^2(t)\ru = 
\lu \langle A\rangle_{\!\rho(t)}^2\ru 
-
\lu \langle A\rangle_{\!\rho(t)}\ru^2 \ ,
\label{125b}
\end{eqnarray}
see equation (\ref{110}).

To evaluate the last term in (\ref{125b}), 
we recast equation (\ref{70}) with (\ref{80}) and 
(\ref{100}) into the form
\begin{eqnarray}
\!\!\!\!
\lu \langle A\rangle_{\!\rho(t)}\ru 
& = & 
F_0(t)\, \langle A\rangle_{\!\rho(0)} 
+
\bar F_0(t) \langle A\rangle_{\!\rhomic}
+ 
R_1(t) 
\label{130b}
\\
R_1(t) & := & \bar F_0(t)\frac{\langle A\rangle_{\!\rhomic}-\langle A\rangle_{\!\rho(0)}}{\DD^2-1}
\label{140b}
\\
\bar{F}_0(t) & := & 1-F_0(t)
\label{150b}
\\
F_0(t) & := & \frac{1}{D^2} \sum_{m,n=1}^{\DD}e^{i (E_n-E_m) t/\hh} = |\phi(t)|^2 \ ,
\label{160b}
\end{eqnarray}
where $\phi(t)$ is defined in equation (\ref{90}).
Similarly as in 
equation (\ref{160}), one sees that
$F_0(t),\,\bar{F}_0(t)\in[0,1]$ for all $t$.
Denoting by $\lmax$ and $\lmin$ the 
largest and smallest among the
eigenvalues $\lambda_1,...,\lambda_{\DD}$ 
of $A$, the range of $A$ is defined as
$\da:=\lmax -\lmin$.
Furthermore, we can and will add a constant 
to $A$ so that $\lmin = -\lmax$
without any change in the final conclusions
below.
It readily follows that $|\lambda_\nu|\leq\da/2$
for all $\nu$ and hence that
\begin{eqnarray}
|\langle A^{\kappa} \rangle_{\!\rho}| \leq (\da/2)^{\kappa}
\label{170b}
\end{eqnarray}
for arbitrary density operators $\rho$
and $\kappa\in{\mathbb N}$.
We thus can infer from equation (\ref{140b}) that
\begin{eqnarray}
|R_1(t)|\leq \da/(\DD^2-1) \ .
\label{180b}
\end{eqnarray}
Likewise, one finds upon squaring equation (\ref{130b}) 
that
\begin{eqnarray}
\!\!\!\!
\!\!\!\!
\lu \langle A\rangle_{\!\rho(t)}\ru^2
& = & 
(
F_0(t)\, \langle A\rangle_{\!\rho(0)}
+
\bar F_0(t) \langle A\rangle_{\!\rhomic}
)^2 
+ R_2(t)
\label{190b}
\\
|R_2(t)| & \leq & 3 \da^2/(\DD^2-1) \ .
\label{200b}
\end{eqnarray}

Turning to the first term on the right 
hand side of (\ref{125b}),
one can infer, 
similarly as in (\ref{30b}), (\ref{40b}),
from (\ref{10}) and (\ref{10b}) that
\begin{eqnarray}
\!\!\!\!
\!\!\!\!
& & \lu \langle A\rangle_{\!\rho(t)}^2\ru 
=
\!\!\!\!
\sum_{a_1,...,a_4}
\!\!e^{i(E_{a_1}-E_{a_2}+E_{a_3}-E_{a_4})t/\hbar}
X_{a_1 ... a_4}
\label{210b}
\\
\!\!\!\!
\!\!\!\!
& & 
X_{a_1 ... a_4}:=\sum_{b_1,...,b_6} 
\rho_{b_1 b_5} \lambda_{b_2} \rho_{b_3 b_6} \lambda_{b_4} 
\nonumber
\\
\!\!\!\!
\!\!\!\!
& & \qquad \qquad 
\times
\lu 
U_{a_1 b_1} \ldots U_{a_4 b_4} 
U_{a_1 \beta_1}^\ast \ldots U_{a_4 \beta_4}^\ast 
\ru \ ,
\label{220b}
\end{eqnarray}
with 
$\beta_1:=b_2$, 
$\beta_2:=b_5$, 
$\beta_3:=b_4$, 
$\beta_4:=b_6$.
Similarly as below equation 
(\ref{40b}) it follows that 
only those summands may be non-zero,
for which $b_1$ and $b_3$ have ``partners'' 
among $\beta_2$ and $\beta_4$,
and vice versa.
This condition can be satisfied in two ways:
(i) $b_5=b_1$ and $b_6=b_3$. 
(ii) $b_5=b_3$ and $b_6=b_1$ and $b_1 \not = b_3$.
The latter condition is due to the fact that
the case $b_1=b_3$ is already covered by (i).
Exploiting (\ref{20b}) and
with the abbreviation $\vec a:=(a_1,...,a_4)$
and likewise for $\vec b$, $\vec \beta$ etc.,
we thus obtain
\begin{eqnarray}
\!\!\!\!
\!\!\!\!
& & X_{\vec a}= X^{(i)}_{\vec a}+X^{(ii)}_{\vec a}
\label{230b}
\\
\!\!\!\!
\!\!\!\!
& & 
X^{(i)}_{\vec a}:=\sum_{\vec b} 
\rho_{b_1 b_1} \lambda_{b_2} \rho_{b_3 b_3} \lambda_{b_4} 
\nonumber
\\
\!\!\!\!
\!\!\!\!
& & \qquad \qquad 
\times
\sum\limits_{P,P'}
V_{P,P'} \prod_{j=1}^4 \delta_{a_j a_{P(j)}} \delta_{b_j\beta^{(i)}_{P'(j)}} 
\label{240b}
\\
\!\!\!\!
\!\!\!\!
& & 
X^{(ii)}_{\vec a}:=\sum_{\vec b,b_1\not=b_3} 
\rho_{b_1 b_3} \lambda_{b_2} \rho_{b_3 b_1} \lambda_{b_4} 
\nonumber
\\
\!\!\!\!
\!\!\!\!
& & \qquad \qquad 
\times
\sum\limits_{P,P'}
V_{P,P'} \prod_{j=1}^4 \delta_{a_j a_{P(j)}} 
\delta_{b_j\beta^{(ii)}_{P'(j)}} 
\ ,
\label{250b}
\end{eqnarray}
where
$\vec \beta^{(i)}:=(b_2,b_1,b_4,b_3)$ 
and
$\vec \beta^{(ii)}:=(b_2,b_3,b_4,b_1)$.

There are $4!=24$ permutations $P$ of the numbers $1,2,3,4$.
Adopting the shorthand notation $[P(1)P(2)P(3)P(4)]$
to explicitly specify a given $P$, these $24$
permutations are:
\begin{eqnarray}
& & 
\!\!\!\!
\!\!\!\!
\!\!\!\!
\!\!\!\!
P_1=[1234]
,\, 
P_2=[2134]
,\, 
P_3=[3214]
,\, 
P_4=[4231],
\nonumber
\\
& & 
\!\!\!\!
\!\!\!\!
\!\!\!\!
\!\!\!\!
P_5=[1324]
,\, 
P_6=[1432]
,\, 
P_7=[1243]
,\, 
P_8=[2143],
\nonumber
\\
& & 
\!\!\!\!
\!\!\!\!
\!\!\!\!
\!\!\!\!
P_9=[3412]
,\, 
P_{10}\!=\![4321]
,\, 
P_{11}\!=\![1342]
,\, 
P_{12}\!=\![1423],
\nonumber
\\
& & 
\!\!\!\!
\!\!\!\!
\!\!\!\!
\!\!\!\!
P_{13}\!=\![3241]
,\, 
P_{14}\!=\![4213]
,\, 
P_{15}\!=\![2431]
,\, 
P_{16}\!=\![4132],
\nonumber
\\
& & 
\!\!\!\!
\!\!\!\!
\!\!\!\!
\!\!\!\!
P_{17}\!=\![2314]
,\, 
P_{18}\!=\![3124]
,\, 
P_{19}\!=\![2341]
,\, 
P_{20}\!=\![2413],
\nonumber
\\
& & 
\!\!\!\!
\!\!\!\!
\!\!\!\!
\!\!\!\!
P_{21}\!=\![3421]
,\, 
P_{22}\!=\![3142]
,\, 
P_{23}\!=\![4312]
,\, 
P_{24}\!=\![4123].
\nonumber
\end{eqnarray}
Observing that
$\beta^{(i)}_j=b_{P_8(j)}$ 
and
$\beta^{(ii)}_j=b_{P_{19}(j)}$ for all $j=1,...,4$,
it is quite straightforward but very arduous
to explicitly carry out the sums over 
$P'$ and $\vec b$ 
in (\ref{240b}), (\ref{250b})
and the sum over $\vec a$ in (\ref{210b}),
yielding
\begin{eqnarray}
\!\!\!\!
\!\!\!\!
& & \lu \langle A\rangle_{\!\rho(t)}^2\ru 
=\sum_{k=1}^{24} f_k(t)\, T(P_k) \ ,
\label{270b}
\end{eqnarray}
where the functions $f_k(t)$ are given by
\begin{eqnarray}
\!\!\!\!
\!\!\!\!
& & 
f_1(t)=D^4F_0^2(t) \ ,
\nonumber
\\
\!\!\!\!
\!\!\!\!
& & 
f_2(t)=f_4(t)=f_5(t)=f_7(t)=D^3F_0(t) \ ,
\nonumber
\\
\!\!\!\!
\!\!\!\!
& & 
f_3(t)=f_6^\ast(t)=D^3 [\phi(t)]^2[\phi(2t)]^\ast \ ,
\nonumber
\\
\!\!\!\!
\!\!\!\!
& & 
f_8(t)=f_{10}(t)=D^2 \ ,
\nonumber
\\
\!\!\!\!
\!\!\!\!
& & 
f_9(t)=D^2F_0(2t) \ ,
\nonumber
\\
\!\!\!\!
\!\!\!\!
& & 
f_k(t)=D^2F_0(t)\ \mbox{for $k=11,...,18$} \ ,
\nonumber
\\
\!\!\!\!
\!\!\!\!
& & 
f_k(t)=D \ \mbox{for $k=19,...,24$} \ ,
\label{280b}
\end{eqnarray}
and the coefficients $T(P)$ are given by
\begin{eqnarray}
\!\!\!\!
\!\!\!\!
& & 
T(P) = 
D^2\langle A\rangle^2_{\!\rhomic}
(
V_{P,P_{8}}+V_{P,P_{24}} \tr\{\rho^2(0)\}
)
\nonumber
\\
\!\!\!\!
\!\!\!\!
& & \ \ 
+
D\langle A^2\rangle_{\!\rhomic}
(
V_{P,P_{10}} \tr\{\rho^2(0)\} +V_{P,P_{19}}
)
\nonumber
\\
\!\!\!\!
\!\!\!\!
& & \ \ 
+ D\langle A\rangle_{\!\rhomic}\langle A\rangle_{\!\rho(0)}
(
V_{P,P_{2}}+V_{P,P_{7}}+V_{P,P_{20}}+V_{P,P_{22}}
)
\nonumber
\\
\!\!\!\!
\!\!\!\!
& & \ \ 
+ D\langle A\rangle_{\!\rhomic}  \tr\{\rho^2(0)A\}
(
V_{P,P_{12}}+V_{P,P_{14}}+V_{P,P_{16}}+V_{P,P_{18}}
)
\nonumber
\\
\!\!\!\!
\!\!\!\!
& & \ \ 
+
\langle A\rangle^2_{\!\rho(0)}
(
V_{P,P_{1}}+V_{P,P_{9}}
)
\nonumber
\\
\!\!\!\!
\!\!\!\!
& & \ \ 
+
\tr\{[\rho(0)A]^2\}
(
V_{P,P_{3}}+V_{P,P_{6}}
)
\nonumber
\\
\!\!\!\!
\!\!\!\!
& & \ \ 
+
\langle A^2\rangle_{\!\rho(0)}
(
V_{P,P_{11}}+V_{P,P_{13}}+V_{P,P_{15}}+V_{P,P_{17}}
)
\nonumber
\\
\!\!\!\!
\!\!\!\!
& & \ \ 
+
\tr\{\rho^2(0)A^2\}
(
V_{P,P_{4}}+V_{P,P_{5}}+V_{P,P_{21}}+V_{P,P_{23}}
)
\ .
\label{290b}
\end{eqnarray}

To explicitly evaluate (\ref{270b})-(\ref{290b}), 
we still need the coefficients $V_{P_k,P_l}$ 
for all $k,l\in\{1,...,24\}$.
They are obtained as explained below equation (\ref{20b}):
Defining $j=j(k,l)$ implicitly via $P_{j}=P_l^{-1}P_k$,
one finds by factorizing each $P_j$ into its 
disjoint cycles and exploiting 
Tables II and IV of Ref. \cite{bro96} 
that $V_{P_k,P_l}$ is given by
\begin{eqnarray}
V_{1,1,1,1} & = & \DD^{-4}\ \ \mbox{for $j=1$},
\nonumber
\\
V_{2,1,1} & = & -\DD^{-5}\ \ \mbox{for $j=2,...,7$},
\nonumber
\\
V_{2,2} & = & \DD^{-6}\ \ \mbox{for $j=8,...,10$},
\nonumber
\\
V_{3,1} & = & 2\, \DD^{-6}\ \ \mbox{for $j=11,...,18$},
\nonumber
\\
V_{4} & = & -5\, \DD^{-7}\ \ \mbox{for $j=19,...,24$},
\label{300b}
\end{eqnarray}
up to correction factors of the form $1+\ord(\DD^{-2})$
on the right hand side of each of those relations.
One thus is left with finding 
$P_{j}=P_l^{-1}P_k$ for all $24^2$ 
pairs $(k,l)$.
To mitigate this daunting task, we have restricted 
ourselves to those summands in (\ref{270b}) which
are at least of the order $D^{-1}$.
Along these lines, one finally recovers with 
equations (\ref{125b}), (\ref{170b}), and (\ref{190b}) 
the result (\ref{120}).

\subsection*{Derivation of equation (\ref{140})}
While the essential steps in deriving equation
(\ref{140}) have been outlined already in 
the main text,
we still have to provide the details 
of the statements below (\ref{140}):
Our first observation is that $R_1(t)$ in 
equation (\ref{130b}) 
amounts to the systematic ($U$-independent) 
part of the omitted corrections in (\ref{140})
and equation (\ref{180b}) to the
bound announced below (\ref{140}).

By means of a straightforward (but again very tedious)
generalization of the calculations from the preceding 
subsection one finds that
\begin{eqnarray}
\lu \xi(t)\xi(s)\ru = 
C(t,s)\frac{\da^2\tr\{\rho^2(0)\}}{\DD} +\ord \left(\frac{\da^2}{\DD^2}\right)
\label{320b}
\end{eqnarray}
where $C(t,s)$ has the 
following six properties: 
First, $C(t,s)=C(s,t)=C(-t,-s)$ for all $t,s$.
Second, $|C(t,s)|\leq 9$ for all $t,s$.
Third, $C(t,0)=0$ for all $t$.
Fourth, $C(t,s) \tto 0$ for $|t-s|\to\infty$, cf. equation (\ref{170}).
Fifth, $C(t,s)\tto F(t-s)
\langle (A-\langle A\rangle_{\!\rhomic})^2 \rangle_{\!\rhomic}$
for $t,s\to\infty$.
Sixth, given $s$, the behavior of $C(t,s)$ as a function of 
$t$ is roughly comparable to that of $F(t-s)$ 
for most $t$.

Though we did not explicitly evaluate 
the last term in (\ref{320b}), closer 
inspection of its general structure 
shows that it can be bounded 
in modulus by $c\, \da^2/\DD^2$ for some $c$ 
which is independent of $t,s,\DD,A,\rho(0),H$.
Moreover, there is no indication of any 
fundamental structural differences in 
comparison with the leading and 
next-to-leading order terms, which 
we did evaluate.
In other words, the last term in (\ref{320b})
is expected to 
satisfy properties analogous 
to those mentioned below equation (\ref{320b}).
Recalling that the purity 
$\tr\{\rho^2(0)\}$ satisfies the usual bounds 
$1\geq\tr\{\rho^2(0)\}\geq\tr\{\rhomic^2\}=1/\DD$, 
we thus recover the properties of $\xi(t)$ 
announced below equation (\ref{140}).

\vspace*{1cm}
{\bf{Acknowledgments}}
\\
I am indebted to Walter Pfeiffer and Thomas Dahm
for numerous enlightening discussions.
I also thank all authors of Refs. \cite{bar09,rig09a,tho04,het15}
for providing the raw data 
of their published works,
in particular Christian Bartsch, Marcos Rigol, 
Tillmann Klamroth, and Daniel Hetterich.
This project was supported by DFG-Grant RE1344/7-1.

\newpage
\section*{SUPPLEMENTARY INFORMATION}

\section*{Supplementary Note 1}
\label{a2}
This note provides a brief account of
those previous analytical findings
which exhibit some appreciable similarity 
to ours.

Ref. [1] (see Supplementary References) 
considers the convergence (for most times)
towards some steady state, 
which is in general different from 
the microcanonical ensemble.
Furthermore, the focus is either
on two-outcome measurements, where one of the 
projectors is of low rank, or the initial state 
must be an eigenvector of the considered observable.
Finally, a role more or less similar to our 
present randomization 
via $U$ is played by the assumption that
the initial state must be spread over
very many energy levels.
Within this setting, 
general upper bounds are obtained
for some suitably defined 
equilibration time scale
(as opposed to the (approximate) 
equality (13) for the entire temporal
behavior).
Apart from these quite significant 
differences, the essential conclusions
are analogous to ours, namely 
an extremely rapid relaxation 
for all above mentioned
two-outcome measurements 
of low rank, as well as
for most observables 
if the initial state is an 
eigenvector of the observable.

Ref. [2] focuses on subsystem-plus-bath
compounds, the total Hilbert space being
a collection of many smaller units
(e.g. due to a local Hamiltonian on a lattice), 
and on separable initial states.
Under these premises, upper bounds
for the subsystem's temporal relaxation
are derived,
which exhibit some (limited) similarities 
to our present findings, including the
prediction of typically very fast
relaxation processes.

Under the additional assumptions that
in the latter setup the 
subsystem is a single 
qubit,  the initial state of the qubit 
as well as the considered observable 
are given by Pauli matrices
(or the identity), 
and the environment is in a pure initial state,
similar findings as in our present work 
have been obtained in Ref. [3].
Note that those initial states of the qubit
are not very physical and that their
linear superposition is not admitted 
in the findings of [3] due to
the non-linearity of the problem.

Refs. [4,5] focus on macroscopic observables 
with a concomitant 
projector $P_{\rm{neq}}$ onto a very small 
subspace of the ``energy shell'' $\hr$ so that any 
(normalized) state $|\psi\rangle\in\hr$  with 
$\langle\psi | P_{\rm{neq}} |\psi\rangle \ll 1$
represents thermal equilibrium.
Denoting, similarly as in our present 
approach, by $U$ the transformation
between the bases of the ``observable''
$P_{\rm{neq}}$ and the Hamiltonian 
$H$, it is then shown that most $U$ 
result in an extremely quick
thermalization for any initial 
pure state $|\psi(0)\rangle\in\hr$.
Similarly as in [1] (see above), 
this conclusion is based on an upper 
(but arguably rather tight) estimate
for the actual temporal relaxation
and on similar assumptions about
the energy level density $\rho(x)$ as in 
equations (17)-(20).

In Ref. [6] it is shown that the 
vast majority of all pure states featuring 
a common expectation value of some generic 
observable at a given time will yield very 
similar expectation values of the same 
observable at any later time.
While in our present approach, $\rho(0)$ and $A$ are
kept fixed relatively to each other and $U$ randomizes 
their constellation relatively to the Hamiltonian $H$,
in Ref. [6] the pair $A$ and $H$ is kept fixed, 
while $\rho(0)$ is randomly sampled under the additional 
constraint that it is a pure state with a preset (arbitrary but fixed)
expectation value $\langle A\rangle_{\!\rho(0)}$.
Moreover, no quantitative statements about how
$\langle A\rangle_{\!\rho(t)}$ actually evolves in time
have been obtained in [6].

Ref. [7] suggests fairly rough 
relaxation time estimates by exploiting
three quite drastic {\em a priori} assumptions.
One of them postulates that the relaxation is
monotonous in time, which can in fact not
be generally true, 
see equation (19) and Fig. 6.
Apart from that, the obtained estimates
are roughly comparable to ours.

\section*{Supplementary Note 2}
\label{a3}
This note compiles some
additional remarks and extensions,
ordered according to their 
appearance in the main text.

\subsubsection*{Regarding section ``Setup''}

1. 
In the present paper, we mainly have in 
mind the examples mentioned below 
equation (1), i.e., $\hr$ represents some
microcanonical ``energy shell''
of a closed many-body system.
But similarly as in Refs. [8-10],
our main result (13) is actually valid for the more 
general setup outlined above equation (1),
i.e., $\hr$ may also represent a more abstract
type of ``active Hilbert space''.
For instance, this may be of 
interest for autonomous systems with 
few degrees of freedom in the context 
of semiclassical chaos when
the initial state is ``spread'' 
over many energy levels.

2. 
{\em A priori}, the pertinent Hilbert 
space of a many-body system is 
not a microcanonical energy 
shell $\hr$, nor are the Hamiltonian, 
observables, and system states given 
by Hermitian operators 
$H$, $A$, and $\rho(t)$ 
on $\hr$ right 
from the beginning.
Rather, the system originally 
``lives'' in a much larger Hilbert 
space $\hr'$ and the Hamiltonian,
observables, and system states
are given by Hermitian operators 
$H'$, $A'$, and $\rho'(t)$ on $\hr'$.
How to go over from the original 
(primed) to the reduced (unprimed) 
setup is not very difficult [11-15], 
but also not entirely obvious:

Similarly as in the main text, 
we denote by $E_n$ and $|n\rangle$
the eigenvalues and eigenvectors of $H'$,
where $n$ runs from $1$ to infinity or to 
some finite upper limit (dimension of $\hr'$). 
Likewise, the corresponding matrix 
elements of $\rho'(t)$ are denoted as 
$\rho'_{mn}(t):=\langle m|\rho'(t)|n\rangle$.
The key point consist in our assumption 
below equation (1) that the system exhibit 
a well defined macroscopic energy, i.e., 
there exists a microcanonical energy window 
$I:=[E-\Delta E,E]$ so that the level 
populations $\rho'_{nn}(0)$ are negligibly 
small for energies $E_n$ outside the 
interval $I$.
Moreover, we can an will assume that 
the labels $n$ and the integer 
$\DD$ are chosen so that 
$E_n\in I \Leftrightarrow n\in\{1,...,\DD\}$.
Next, we denote by $\hr$ the subspace 
spanned by $\{|n\rangle\}_{n=1}^\DD$, by
$P:=\sum_{n=1}^\DD|n\rangle\langle n|$
the projector onto $\hr$,
and by $H:=PH'P$, $A:=PA'P$, $\rho(t):=P\rho'(t)P$ 
the corresponding ``restrictions'' or 
``projections'' of the original operators.
With $|\rho'_{mn}|^2\leq \rho'_{mm} \rho'_{nn}$
(Cauchy-Schwarz inequality)
and the above approximation $\rho'_{nn}(0)=0$
for $n>\DD$, it follows that $\rho'_{mn}(0)=0$ 
if $m>\DD $ or $n>\DD $ and hence
that $\rho(0)=\rho'(0)$.
Since $P$ commutes with $H'$
and thus with $\propb:=e^{-iH't/\hh}$,
the original time evolution
$\rho'(t)=\propb\rho'(0)(\propb)^\dagger$
implies that $\rho(t)=\rho'(t)$
for all $t$, and with $P^2=P$ it 
follows that $\rho(t)=\propa\rho(0)\propa^\dagger$,
where $\propa:=e^{-iHt/\hh}$.
Exploiting the cyclic invariance of the 
trace and $P^2=P$ finally yields 
$\tr\{\rho'(t) A'\}=\tr\{\rho(t) A\}$
for all $t$.

So far, the basic operators $H$, $A$, 
$\rho(t)$ and their descendants 
$\propa$ and $\tr\{\rho(t) A\}$ are 
strictly speaking still defined on 
$\hr'$ but it is trivial to 
reinterpret them as being defined 
on $\hr$.
In particular, the eigenvalues and 
eigenvectors of $H : \hr \to \hr$ are
now given by $\{E_n\}_{n=1}^\DD$ 
and $\{|n\rangle \}_{n=1}^\DD$, respectively.
While the connection between
$H$ and $H'$ and between $\rho(t)$
and $\rho'(t)$ is thus rather trivial,
the eigenvalues and eigenvectors of 
$A$ are in general quite different from 
those of $A'$.
Nevertheless, all original (primed)
expectation values are correctly 
recovered within the reduced 
(unprimed) formalism.

\subsubsection*{Regarding section ``Analytical results''}

3.
A natural intuitive guess is that
$\rho_{nn}(0)$ and $A_{nn}$ 
should be essentially independent
of each other in the sense that
$\lu\rho_{nn}(0)A_{nn}\ru$ 
can be approximated by
$\lu\rho_{nn}(0)\ru\lu A_{nn}\ru$.
If so, one could readily conclude from equations (2) and (4)
that $\langle A\rangle_{\!\rhou}=\langle A\rangle_{\!\rhomic}$,
which is nothing else than the leading order
approximation of equation (9).
In other words, our guess seems right,
the essence of equation (9) is 
intuitively quite obvious,
and the last term in equation (9) 
must be due to weak correlations between 
$\rho_{nn}(0)$ and $A_{nn}$ for 
non-equilibrium initial conditions
$\langle A\rangle_{\!\rho(0)}$.

\subsubsection*{Regarding section ``Basic properties of $F(t)$''}

4.
The essential prerequisite in approximating 
equation (8) by (17) is that 
$t/\hbar$ must be much smaller than the
inverse mean level distance.
Since the energy levels are extremely 
dense for typical many-body systems,
the approximation applies for
all experimentally realistic times $t$.
However, the quasi-periodicities of $F(t)$
for extremely large $t$, 
inherited from $\phi(t)$ via (7) and (8),
usually get lost. 

\subsubsection*{Regarding section ``Typicality of thermalization''}

5.
By similar methods as in the derivation of 
our main result (13), one can show [11]
that for the overwhelming majority of unitaries $U$ 
the diagonal matrix elements $A_{nn}$ remain very 
close to their mean value
$\lu A_{nn}\ru=\langle A\rangle_{\!\rhomic}$,
a property also known under the name eigenstate 
thermalization hypothesis (ETH) [12-14].
It is tempting to argue that a violation of
ETH indicates an ``untypical'' case and hence 
also (13) will be violated.
However, there is 
no reason why the extremely small subset 
of $U$'s which violate (13) has any 
relevant overlap with the extremely small 
subset of $U$'s which violate ETH.
In other words, we expect that 
equation (13) still applies to
the vast majority of ETH-violating systems,
i.e., provided their initial condition
$\rho(0)$ is still sufficiently ``typical''
to guarantee thermalization.
Numerical examples of such cases are 
provided, e.g., by Ref. [15].

Vice versa, the findings about typicality
of ETH and thermalization from
[11,12,16-19]
are expected to remain valid 
even when (13) is violated
(thus including cases which do not
thermalize as rapidly as
predicted by (13)).

An analogous consideration applies to the
``level populations'' $\rho_{nn}(0)$:
They must be negligible outside the
microcanonical energy window $[E-\Delta E,E]$,
but inside the window they may still 
be distributed quite ``untypically''.

6.
More abstractly speaking, in order to realize 
simultaneously an untypical $U$ and a far 
from equilibrium $\langle A\rangle_{\!\rho(0)}$, 
one generally expects that the eigenbases of 
both $A$ and $\rho(0)$ must be 
fine-tuned relatively to a given $H$.

As a consequence, one expects untypically strong 
correlations between $A_{nn}$ and $\rho_{nn}(0)$
(see also paragraph 3. above).
This is confirmed, e.g., 
by the numerical examples in 
Refs. [14,20] and is also
closely related to the ideas proposed
by Peres in Ref. [21].

7.
To further scrutinize the untypical $U$'s,
we consider subsets $S_a$ consisting of all $U$'s with 
the extra property that $\langle A\rangle_{\! \rhobar}=a$,
where $\rhobar$ is defined below equation (3).
One readily sees that for any given $a$-value, 
the set $S_a$ still entails the necessary 
symmetries so that equations 
(2)-(8) remain valid when 
re-defining $\lu \cdots\ru$
as the restricted average over
all $U\in S_a$.
On the other hand, equation (9) is 
replaced by $\langle A\rangle_{\! \rhou}=a$,
implied by $\langle A\rangle_{\! \rhobar}=a$
for all $U\in S_a$.
Finally, one expects, analogously as 
in equations (10)-(12), that the 
fluctuations about the average behavior 
are typically small for most $U \in S_a$.
While a rigorous proof seems very difficult,
the intuitive argument is that the subset $S_a$
can be represented as a manifold
of fantastically large dimensionality
(just one dimension less that for the
unrestricted set of all $U$'s due to 
the extra constraint 
$\langle A\rangle_{\! \rhobar}=a$).
Hence, a similar concentration of 
measure phenomenon is 
expected in both cases.
Analogously as in (13),
the overall conclusion is that
\begin{eqnarray}
\langle A\rangle_{\!\rho(t)}
& = & 
a
+
F(t)\, \left\{\langle A\rangle_{\!\rho(0)} 
- 
a
\right\}
\nonumber
\end{eqnarray}
should be satisfied in very good approximation
for the vast majority of all times $t$ and 
unitaries $U\in S_a$.
Upon comparison with (13) one sees that
if $a$ notably differs from $\langle A\rangle_{\!\rhomic}$ 
then most $U\in S_a$ are untypical.
On the other hand, any given untypical $U$ 
is contained in one of the subsets 
$S_a$ and is thus generically expected 
to satisfy the above approximation.
Remarkably, the time dependence 
is governed by the same function $F(t)$ 
for all $a$.

These considerations justify our comparison 
of the theory (in the above generalized 
version) with the integrable model in Fig. 4.
In turn, the good agreement
with the numerical results in Fig. 4 
supports the above arguments.

\subsection*{Supplementary References}
\begin{description}

\item[ ]
[1]
Malabarba, A. S. L., Garcia-Pintos, L. P., Linden, N., 
Farrelly, T. C., \& Short, A. J. 
Quantum systems equilibrate rapidly for most observables.
{\em Phys. Rev. E} {\bf 90}, 012121 (2014)

\item[ ]
[2]
Cramer, M. 
Thermalization under randomized local Hamiltonians.
{\em New. J. Phys.} {\bf 14}, 053051 (2012)

\item[ ]
[3]
Znidaric, M., Pineda, C., \& Garcia-Mata, I.
Non-Markovian behavior of small and large complex quantum systems.
{\em Phys. Rev. Lett.} {\bf 107}, 080404 (2011)

\item[ ]
[4]
Goldstein, S., Hara, T., \& Tasaki, H.
Extremely quick thermalization in a macroscopic quantum system for a typical nonequilibrium subspace.
{\em New. J. Phys.} {\bf 17}, 045002 (2015)

\item[ ]
[5]
Goldstein, S., Hara, T., \& Tasaki, H.
The approach to equilibrium in a macroscopic quantum 
system for a typical nonequilibrium subspace.
Preprint at http://arxiv.org/abs/1402.3380
(2014)

\item[ ]
[6]
Bartsch, C. \& Gemmer, J. 
Dynamical typicality of quantum expectation values.
{\em Phys. Rev. Lett.} {\bf 102}, 110403 (2009)

\item[ ]
[7]
Monnai, T. 
Generic evaluation of relaxation time for quantum many body systems:
analysis of system size dependence.
{\em J. Phys. Soc. Jpn.} {\bf 82}, 044006 (2013)

\item[ ]
[8]
Popescu, S., Short, A. J., \& Winter, A. 
Entanglement and the foundations of statistical mechanics.
{\em Nature Phys.} {\bf 2}, 754-758 (2006)

\item[ ]
[9]
Popescu, S., Short, A. J., \& Winter, A.
The foundations of statistical mechanics from entanglement: 
Individual states vs. averages.
Preprint at
http://arxiv.org/abs/quant-ph/0511225
(2005)

\item[ ]
[10]
Linden, N., Popescu, S., Short, A. J., \& Winter, A. 
Quantum mechanical evolution towards equilibrium.
{\em Phys. Rev.  E} {\bf 79}, 061103 (2009)

\item[ ]
[11]
von Neumann, J.
Beweis des Ergodensatzes und des H-Theorems 
in der neuen Mechanik.
{\em Z. Phys.} {\bf 57}, 30-70 (1929)
$[$English translation by Tumulka, R.
Proof of the ergodic theorem and the H-theorem 
in quantum mechanics.
{\em Eur. Phys. J. H} {\bf 35}, 201-237 (2010)$]$

\item[ ]
[12]
Goldstein, S., Lebowitz, J. L., Tumulka, R., \& Zhang\`{\i}, N.
Long-time behavior of macroscopic quantum systems: commentary accompanying 
the english translation of John von Neumann's 1929 article on the quantum ergodic theorem.
{\em Eur. Phys. J. H} {\bf 35}, 173-200 (2010)

\item[ ]
[13]
Goldstein, S., Lebowitz, J. L., 
Mastrodonato, C., Tumulka, R., \& Zhang\`{\i}, N.
Approach to thermal equilibrium of macroscopic quantum systems.
{\em Phys. Rev.  E} {\bf 81}, 011109 (2010)

\item[ ]
[14]
Goldstein, S., Lebowitz, J. L., 
Mastrodonato, C., Tumulka, R., \& Zhang\`{\i}, N.
Normal typicality and von Neumann's quantum ergodic theorem.
{\em Proc. R. Soc. A} {\bf 466}, 3203-3224 (2010)

\item[ ]
[15]
Reimann, P.
Generalization of von Neumann's approach to thermalization.
{\em Phys. Rev. Lett.} {\bf 115}, 010403 (2015) 

\item[ ]
[16]
Deutsch, J. M. 
Quantum statistical mechanics in a closed system.
{\em Phys. Rev. A} {\bf 43}, 2046-2049 (1991)

\item[ ]
[17]
Srednicki, M. 
Chaos and quantum thermalization.
{\em Phys. Rev. E} {\bf 50}, 888-901 (1994)

\item[ ]
[18]
Rigol, M., Dunjko, V., \& Olshanii, M.
Thermalization and its mechanism for generic
isolated quantum systems.
{\em Nature} {\bf 452}, 854-858 (2008)

\item[ ]
[19]
Rigol, M. \& Srednicki, M. 
Alternatives to eigenstate thermalization.
{\em Phys. Rev. Lett.} {\bf 12}, 110601 (2012)

\item[ ]
[20]
Gogolin, C., M\"uller, M., \& Eisert, J. 
Absence of thermalization in nonintegrable systems.
{\em Phys. Rev. Lett.} {\bf 106}, 040401 (2011)

\item[ ]
[21]
Peres, A.
Ergodicity and mixing in quantum theory I.
{\em Phys. Rev. A} {\bf 30}, 504-509 (1984)

\end{description}

\end{document}